\documentclass[11pt,a4paper]{article}
\usepackage[T1]{fontenc}
\usepackage[utf8]{inputenc}
\usepackage[margin=25mm]{geometry}
\usepackage{mathptmx}
\usepackage{microtype}
\usepackage[super,comma,sort&compress]{natbib}
\usepackage{titlesec}
\usepackage{graphicx}
\usepackage{float}
\usepackage{xcolor}
\usepackage{needspace}
\makeatletter
\NewCommandCopy\arxivKernelCaption\caption
\let\arxiv@kernel@caption\@caption
\let\arxiv@kernel@makecaption\@makecaption
\makeatother
\usepackage{booktabs,longtable,array,caption,amsmath}
\usepackage{listings,textcomp}
\lstdefinestyle{s1code}{basicstyle=\ttfamily\fontsize{8}{10}\selectfont,
  breaklines=true,breakatwhitespace=false,columns=fullflexible,keepspaces=true,
  showstringspaces=false,frame=single,framerule=0.3pt,rulecolor=\color{black!35},
  xleftmargin=3pt,xrightmargin=3pt,framesep=4pt,aboveskip=8pt,belowskip=8pt,
  tabsize=4,upquote=true}
\AtBeginDocument{}

\usepackage[most]{tcolorbox}
\usepackage{pdflscape}
\definecolor{s1promptblue}{RGB}{20,38,168}
\newtcolorbox{s1promptbox}[1]{enhanced,breakable,
  colback=white,colframe=s1promptblue,boxrule=0.7pt,arc=1.5mm,
  left=3mm,right=3mm,top=3mm,bottom=3mm,
  before skip=12pt,after skip=12pt,
  fontupper=\normalfont\fontsize{10}{12}\selectfont,
  title={#1},fonttitle=\bfseries\fontsize{10}{12}\selectfont,
  coltitle=white,colbacktitle=s1promptblue,
  attach boxed title to top center={yshift=-2mm},
  boxed title style={boxrule=0pt,arc=0.8mm,left=2mm,right=2mm,top=1mm,bottom=1mm},
  before upper={\setlength{\parskip}{3pt}\setlength{\parindent}{0pt}},
  pad at break*=2mm}
\lstdefinestyle{s1promptpayload}{basicstyle=\ttfamily\fontsize{8.8}{10.6}\selectfont,
  breaklines=true,breakatwhitespace=false,columns=fullflexible,keepspaces=true,
  showstringspaces=false,frame=none,numbers=none,backgroundcolor=\color{white},
  xleftmargin=0pt,xrightmargin=0pt,aboveskip=4pt,belowskip=2pt,
  tabsize=2,upquote=true}
\usepackage{hyperref}
\hypersetup{hidelinks}

\titleformat{\section}{\normalfont\large\bfseries}{}{0pt}{}
\titlespacing*{\section}{0pt}{1.1\baselineskip}{0.4\baselineskip}
\renewcommand{\bibsection}{\section*{References}}

\hypersetup{pdftitle={Agentic AI for public health data collection}}
\makeatletter\renewcommand*\l@subsection{\@dottedtocline{2}{1.5em}{3.1em}}\makeatother

\begin{document}
\makeatletter
\NewCommandCopy\arxivSIcaption\caption
\let\arxiv@SI@caption\@caption
\let\arxiv@SI@makecaption\@makecaption
\RenewCommandCopy\caption\arxivKernelCaption
\let\@caption\arxiv@kernel@caption
\let\@makecaption\arxiv@kernel@makecaption
\makeatother
{\Large\bfseries Toward an Agentic AI Approach to Public Health Data Collection \par}
\vspace{0.5\baselineskip}

{
Xueyi Zhang\textsuperscript{a},
Hongru Du\textsuperscript{a,*},
Ensheng Dong\textsuperscript{b,*}
\par}
\vspace{0.3\baselineskip}

{\small
\textsuperscript{a}Department of Systems and Information Engineering, University of Virginia, Charlottesville, USA\\
\textsuperscript{b}Engineering and Applied Sciences Department, RAND, Arlington, USA\\
\textsuperscript{*}Corresponding author: \href{mailto:hongrudu@virginia.edu}{hongrudu@virginia.edu},
\href{mailto:edong@rand.org}{edong@rand.org}
\par}

\begin{abstract}
Disease surveillance increasingly integrates automated data collection, yet predefined sources and procedures require manual adaptation as reporting practices change. We explore agentic artificial intelligence (AI) as an approach to coordinating surveillance data collection across heterogeneous sources. Specifically, we built a multi-agent framework that integrates source discovery, source assessment, evidence structuring, event linking, and validation. Three retrospective applications illustrate its promise. The workflow recovered all 13 individual cases in a hantavirus outbreak, with links to 85 distinct publisher domains compared with 28 in the reference benchmark, reconstructed Canadian measles counts with a final total 0.33\% below the national reference and higher cumulative counts on 79.8\% of comparable days, and increased the share of days with recent Sierra Leone mpox reports from 28.2\% to 65.6\% by incorporating attributed media. These demonstrations also reveal limitations in record completeness and accuracy, while prospective gains in timeliness remain untested. We envision agentic AI as a complementary layer of public health data infrastructure, coordinating evidence-linked collection under surveillance goals and standards defined by public health professionals.

\end{abstract}

\begin{figure}[h!]
\centering
\IfFileExists{Figure_1.pdf}{\includegraphics[width=\textwidth]{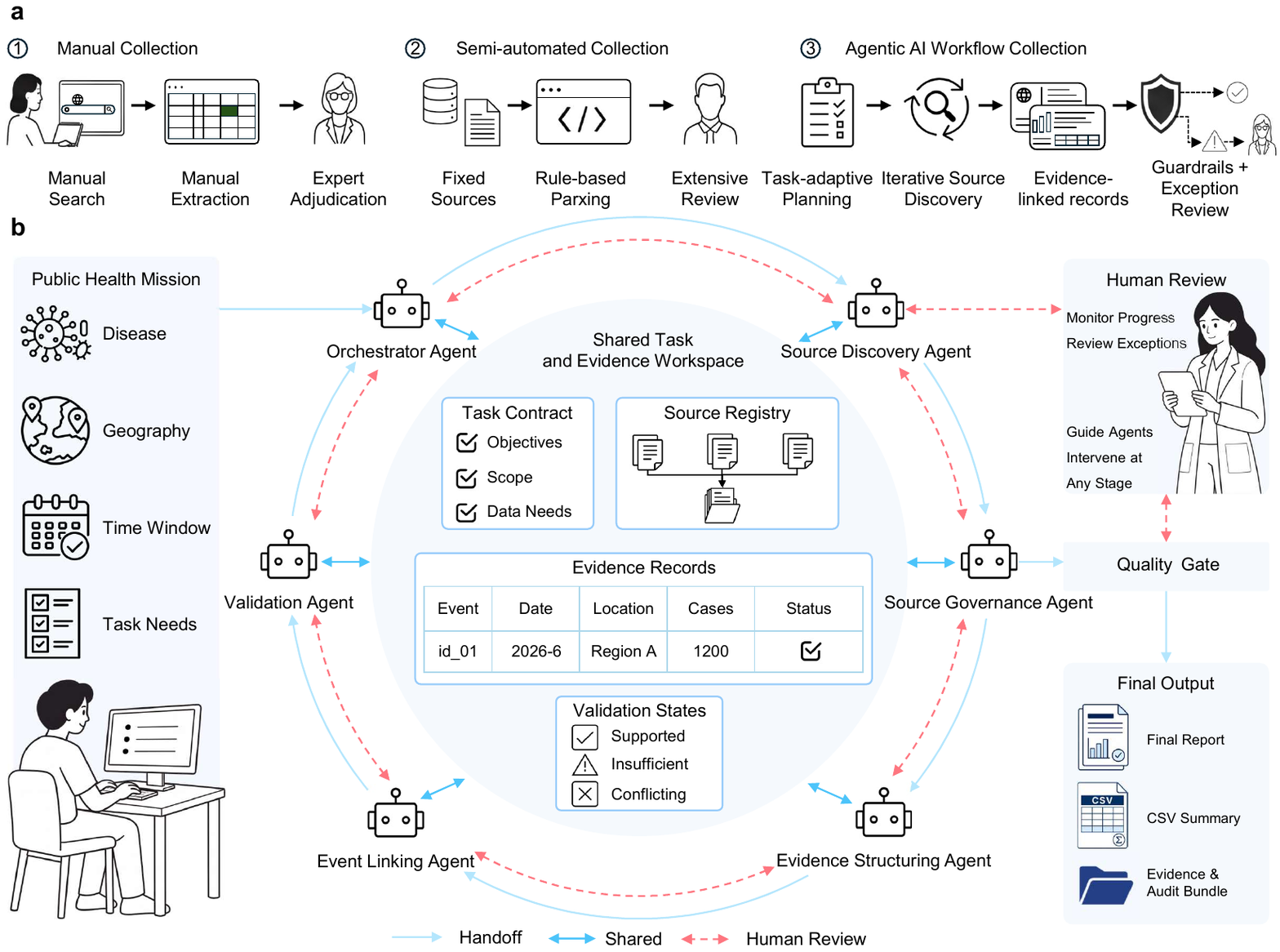}}{\includegraphics[width=\textwidth]{figure 1.pdf}}
\caption{\textbf{Evolution of public health surveillance and the proposed agentic AI workflow.} \textbf{a}, Conceptual evolution of public health data surveillance from manual, in which collections are performed by humans, to semi-automated systems built around fixed sources and rule-based processing, and toward agentic AI workflows combining end-to-end automation, adaptive execution, and expert oversight. \textbf{b}, Proposed agentic AI paradigm for public health data surveillance. A public health mission is translated into a shared task contract that coordinates specialized agents for source discovery, source assessment, evidence structuring, event linking, and validation. Gaps and conflicts redirect the workflow or trigger expert review, and a final quality gate releases structured outputs with traceable evidence. }
\label{fig1}
\end{figure}

Disease surveillance is essential to the prevention and control of infectious diseases~\cite{cdc,langmuir1971}, and the collection of outbreak information has evolved from manual investigation to increasingly automated systems~\cite{lee2011} (Fig.~\ref{fig1}a). Early surveillance relied largely on manual search and extraction, pioneered by John Snow’s investigation of London's 1854 cholera outbreak~\cite{snow}. Digital surveillance systems subsequently expanded the scale and speed of information collection, while platforms such as GPHIN~\cite{gphin}, EpiSPIDER~\cite{epispider}, HealthMap~\cite{healthmap} and, later, the Johns Hopkins CSSE COVID-19 dashboard~\cite{dong2020} increasingly combined automated collection and processing with expert review. Despite these advances, many surveillance systems continue to rely on predefined sources and source-specific processing procedures, requiring substantial human adaptation as reporting environments evolve~\cite{dong2022}.

The challenge extends beyond automating information retrieval and extraction. Surveillance data appear in diverse formats, including narrative text, tables, and scanned documents \cite{dong2022,timeliness, harrod}, while differences in case definitions, reporting periods, and geographical units complicate the integration of information across sources \cite{dong2022,pandem,kraemer2025}. Distinguishing new from cumulative case counts, for example, requires preserving the meaning of the source report during automated processing~\cite{sentinel}. Ensuring consistency therefore requires linking extracted observations to their supporting evidence \cite{xu,mpox}, allowing discrepancies to be identified and examined across reports~\cite{hed,eioskorea}. Ambiguous or conflicting information may still require verification with relevant authorities and interpretation by public health experts \cite{epicore,eiosafrica}. Together, these challenges highlight the need to move beyond automating individual collection tasks toward workflows that can adapt to changing information environments, preserve evidence traceability, and direct expert attention to decisions requiring epidemiological judgment.

Agentic AI offers a potential pathway toward more adaptive public health data collection. Rather than executing predefined processing steps, agentic AI systems pursue specified goals through iterative planning, tool use, and feedback, allowing information obtained during execution to shape subsequent actions~\cite{toolformer,react}. Persistent task state and coordination across specialized agents further support adaptation across interconnected tasks as evidence and requirements evolve~\cite{reflexion,autogen}. Recent studies have used large language models (LLMs) to structure epidemiological information from outbreak reports and proposed agentic frameworks for epidemic intelligence and decision support~\cite{consoli2025,yang2026,bhadelia2026}. Building on these advances, we explore agentic AI as an approach to coordinating public health data collection across heterogeneous sources, shifting from human-maintained, source-specific pipelines toward adaptive agentic workflows guided by expert-defined surveillance goals and evidence standards. Such workflows could reduce the need for repeated manual adaptation while enabling more responsive, evidence-driven data collection with continued expert oversight.

To explore this direction, we developed an agentic workflow for disease surveillance data collection (Fig.~\ref{fig1}b). Users define the surveillance target by specifying the disease, geographical scope, observation period, and data requirements; an orchestrator agent then translates it into a shared task specification. Specialized agents then coordinate source discovery, source governance, evidence structuring, event linking, and validation through a shared task and evidence workspace. The workspace maintains a source registry, evidence-linked observations, and validation states, allowing information gaps and conflicting evidence to redirect subsequent collection and verification actions.  A final quality gate determines which records qualify for release, producing structured surveillance datasets accompanied by supporting evidence, unresolved issues, and an audit trail. See full pipeline implementation details in Supplementary Section S1. 

We applied the designed Agentic AI workflow to three surveillance settings: 1) individual case collection for a 2026 hantavirus outbreak, 2) population-level reconstruction of 2025 measles activity in Canada, and 3) population-level reconstruction of 2025 mpox reporting in Sierra Leone. For hantavirus, the reconstructed case records were compared with an expert-curated benchmark~\cite{globalhealth2026}. For Canadian measles, reports from the Public Health Agency of Canada  (PHAC) were excluded from the reconstructed series and used as an external reference. World Health Organization (WHO) reports served the same role for Sierra Leone mpox.

\begin{figure}[h!]
\centering
\IfFileExists{Figure_2.pdf}{\includegraphics[width=\textwidth]{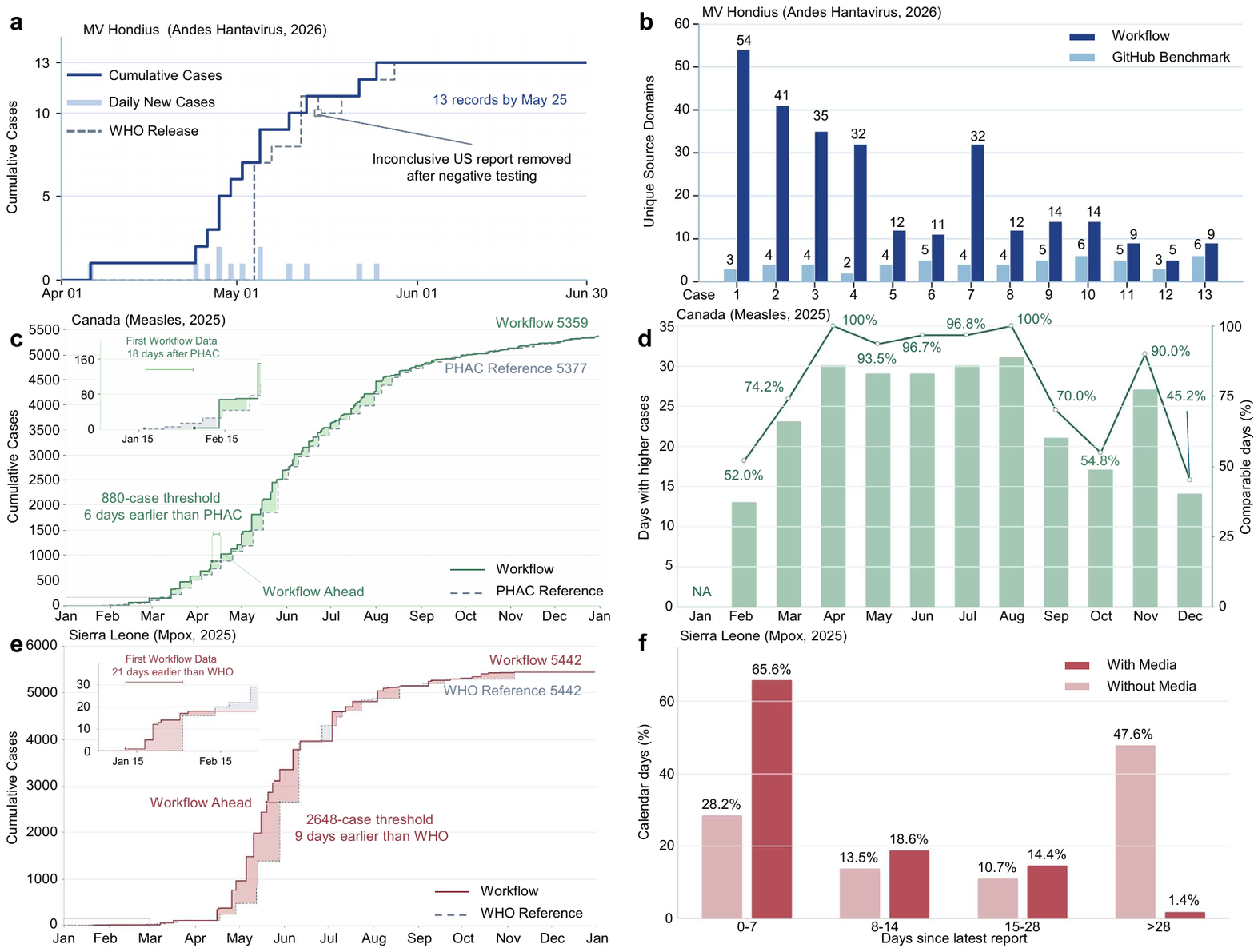}}{\includegraphics[width=\textwidth]{figure 2.pdf}}
\caption{\textbf{Workflow-reconstructed results and reference comparisons across three case studies.}
\textbf{a}, Hantavirus cases (2026): workflow-reconstructed cumulative cases compared with World Health Organization (WHO) reference counts by release date.
\textbf{b}, Distinct case-linked publisher domains per case for the workflow-reconstructed results in dark blue and the GitHub benchmark in light blue.
\textbf{c}, Canadian measles (2025): cumulative reported cases by release date for the workflow-reconstructed results and the Public Health Agency of Canada (PHAC) reference.
\textbf{d}, Number of days per month with workflow cumulative counts above PHAC counts, shown by light-green bars on the left axis, and their percentage among days with both series available, shown by the dark-green line on the right axis. NA indicates no comparable days.
\textbf{e}, Sierra Leone mpox (2025): cumulative confirmed cases by release date for the workflow-reconstructed results and the WHO reference. 
\textbf{f}, Percentage of calendar days by time since the latest available report in 2025, without media in light red and with media in dark red.}
\label{fig2}
\end{figure}

For the 2026 hantavirus outbreak, the workflow reconstructed all 13 confirmed or probable cases represented in the expert-curated benchmark~\cite{globalhealth2026}, with cumulative counts shown alongside WHO updates~\cite{whohantavirus2026} in Fig.~\ref{fig2}a. The reconstructed records linked to 85 distinct publisher domains, compared with 28 in the benchmark, with greater source coverage for every case (Fig.~\ref{fig2}b). The workflow correctly recovered 59 of 87 evaluable fields (67.8\%), compared with 74 (85.1\%) in the manually curated benchmark. Case status was correct for all 13 cases, whereas age-group and symptom-onset information remained less complete (Supplementary Figs. S3–S4 and Table S3). This expanded the supporting evidence available for case-level review, although broader source coverage does not necessarily indicate greater accuracy. Additional results are provided in Supplementary Section S2.

For 2025 Canadian measles, the workflow assembled 217 source observations covering all 13 provinces and territories into a national time series ending at 5,359 cases, just 0.33\% below the PHAC reference~\cite{phac_dec2025} total of 5,377 (Fig.~\ref{fig2}c). Across 47 comparable reference dates, the weighted absolute percentage difference was 1.73\%, with provincial comparisons ranging from 1.97\% in Manitoba to 3.27\% in British Columbia (Supplementary Section S3). Workflow cumulative counts were higher on 264 of 331 comparable days, or 79.8\%, with monthly proportions ranging from 45.2\% to 100\% (Fig.~\ref{fig2}d). These findings illustrate the potential of integrating decentralized reports to complement national surveillance, although higher reported counts do not necessarily indicate greater accuracy.

For Sierra Leone mpox in 2025, the workflow reconstructed a national time series that matched the WHO reference total of 5,442 cases (Fig.~\ref{fig2}e). Of 44 retained source observations, 27 came from attributed media reports. Adding attributed media reports to national official and regional institutional reports increased the proportion of days with a report no more than seven days old from 28.2\% to 65.6\% over the 355-day period (Fig.~\ref{fig2}f). At 24 WHO reference dates, incorporating media reduced the mean absolute count difference from 1,375 to 163 cases, improving agreement on 75\% of reference dates (Supplementary Fig. S11). This demonstrates how integrating heterogeneous reporting sources can improve the continuity of available surveillance information.

These retrospective applications show the potential of moving public health data collection toward adaptive, evidence-linked workflows. A common architecture supported individual-case reconstruction, integration of geographically distributed reports, and more continuous coverage of available outbreak information. The proposed shift is from automating individual tasks to coordinating collection and verification through a shared evidence base, allowing unresolved gaps and conflicts to guide subsequent actions. Agents perform repetitive collection and preliminary checks, while public health professionals define information needs, establish evidence standards, and adjudicate unresolved findings. These workflows could serve as an adaptive layer within existing public health data infrastructure, potentially reducing reliance on source-specific pipelines during emerging outbreaks. Local publications and attributed media could complement delayed or sparse official reporting, while evidence-linked records could support interagency sharing, forecasting, and risk assessment~\cite{bhadelia2026}.

Whether these benefits generalize across surveillance settings and offer measurable advantages over existing approaches remains to be established. Comparisons with withheld official sources assess population-level agreement but do not independently establish accuracy for individual records. Similarly, retrospective reporting leads reflect historical information availability rather than gains in prospective detection. Future evaluations should compare agentic workflows with simpler automated approaches and general-purpose AI tools, assessing record accuracy, time to validated release, computational cost, and human effort. Moreover, evidence traceability supports auditing but does not ensure correctness. Errors can propagate into structured records, underscoring the need for cross-source verification and expert review. Changing webpages, variable model responses, and model updates complicate reproducibility, while robustness across models remains untested in our evaluation. Broader geographical coverage, longer observation periods, and more detailed records may also increase computational and review demands. Agentic collection should therefore complement established reporting systems and expert judgment, with its value assessed through the accuracy and timeliness of released records and the effort required to produce them.

\section*{Methods}

\textbf{Agentic AI surveillance workflow.} We developed an agentic AI workflow to collect infectious disease data from publicly available reports, with each observation linked to its supporting evidence. Each task specified the disease, geographical scope, observation period, and required output structure. In this framework, an orchestrator agent coordinates source discovery, source assessment, evidence structuring, event linking, and validation through a shared workspace (Fig.~\ref{fig1}b; Supplementary Fig. S1 and Table S1). The workflow is implemented in LangGraph, which connects the stages as nodes, maintains a shared state, and selects the next operation through fixed or conditional edges (Supplementary Fig. S2). All case studies used Claude Sonnet 4.6 as the language model. The shared state retains task requirements, sources, documents, evidence fragments, and record assessments to inform later decisions. Targeted search, retrieval, and extraction continue until no eligible action remains or the execution budget is exhausted. Although human review was not used in three case studies, it can be incorporated at any stage to inspect evidence, resolve uncertainties, or guide subsequent actions. All prompts are given in full in Supplementary Section S1.

\textbf{Source discovery.} The task and disease profile guide search queries across relevant source categories and languages. Queries are checked against the task before execution through Tavily Search~\cite{tavily}. Search results are recorded and deduplicated while retaining their discovery history. After each batch, the workflow reviews the result metadata and refines subsequent queries until predefined stopping conditions are met. Further details are provided in Supplementary Section S1.2.

\textbf{Source assessment.} The workflow identifies report publishers and reused material, distinguishing publishers from hosting sites and search providers. A screening step and a source critic assess task relevance and whether sources are likely to contain usable observations, each providing a confidence score. Irrelevant sources are excluded. Assessment procedures and scoring details are provided in Supplementary Section S1.3.

\textbf{Evidence acquisition.} The workflow retrieves reports and relevant attachments from the selected sources and prepares their contents for extraction. Information presented in webpages, documents and tables is organized while preserving the context needed to interpret it. For example, table values remain linked to their headings and explanatory notes. Each piece of evidence is linked to its source, with the retrieval time recorded, so that later observations can be checked against the original material. Further details are provided in Supplementary Section S1.3.

\textbf{Structured extraction and normalization.} The workflow uses language models and predefined rules to organize explicitly reported information into records with a common structure. Missing values are left unfilled, unreported quantities are not calculated, and aggregate counts are not converted into individual cases. Case classifications, count types and date meanings are kept distinct. Extracted values are checked against supporting quotations and their context, and numeric and date formats are standardized while retaining the original values and evidence. Records are classified as valid, needing review or rejected. Further details are provided in Supplementary Section S1.4.

\textbf{Verification across sources.} The workflow links related records using disease, location, timing and count definitions. Records are treated as duplicates only when their counts match and at least one record specifies what the counts represent. One representative is then retained, with the others linked as supporting evidence to avoid double counting. Comparisons distinguish agreement, partial support, repeated use of the same source and conflicting reports. Differences guide further review without automatically changing reported values. Records are checked against their supporting evidence before inclusion in the dataset. Incomplete or conflicting information can prompt further searches. Further details are provided in Supplementary Sections S1.5 and S1.6.

\textbf{Experimental settings.} We evaluated individual-case reconstruction for the MV Hondius hantavirus outbreak from 1 April to 30 June 2026, and population-level reconstruction of measles in Canada and mpox in Sierra Leone during 2025. Designated reference datasets and reports were withheld from reconstruction and reserved for external evaluation, with source-use restrictions specified for each application. For hantavirus, the workflow's results were compared with the manually curated Global.health line list~\cite{globalhealth2026}, and clinical fields in both datasets were assessed against the same reference values compiled from the published case table in \textit{Eurosurveillance} (including its 9 July 2026 addendum) and the WHO rapid risk assessment (Global v.2)~\cite{si-eurosurveillance,si-whorra}. For Canadian measles, PHAC reports and reports explicitly attributing national totals to PHAC were excluded from the workflow series and used as the reference. For Sierra Leone mpox, the workflow series combined national official publications, attributed media reports and regional institutional reports. Source assessment considered publisher identity and upstream attribution, although unattributed reuse of reference figures could not be ruled out. WHO publications and institutional reports explicitly attributing their counts to WHO were excluded from this series and used as the reference. Details of the datasets, clinical reference and additional comparisons are provided in Supplementary Sections S2--S4.

\textbf{Evaluation measures.} For hantavirus, we compared case coverage and distinct publisher domains overall and per case against the Global.health benchmark. Clinical fields were evaluated for completeness and agreement with the common reference. Detailed source and field evaluations are provided in Supplementary Section S2. For both population-level series, we used report publication dates. We also compared when each series first reached selected count thresholds. For Canadian measles, the absolute difference between final counts was expressed as a percentage of the PHAC total. We calculated monthly and annual numbers and percentages of days when workflow cumulative counts exceeded PHAC counts, including only days with counts available in both series. For mpox, report age was the number of days since the latest retained report. We compared the percentages of days with report ages of 0--7, 8--14, 15--28 and more than 28 days over the same 355 days from 11 January to 31 December 2025. Further details are provided in Supplementary Sections S3 and S4.

\textbf{Data availability.}  The reconstructed surveillance datasets for the three case studies, together with source URLs and supporting information, are available at \url{https://github.com/XueyiZhang75/data-collection-workflow/tree/main/data/case_studies}. All underlying outbreak information was collected from publicly available reports. The reference data were obtained from Global.health, the PHAC, and the WHO.

\textbf{Code availability.} The data collection workflow code is available at \url{https://github.com/XueyiZhang75/data-collection-workflow}.

\clearpage
\newgeometry{margin=23mm,headheight=14pt}
\savegeometry{si}
\appendix
\setcounter{section}{0}
\setcounter{figure}{0}
\setcounter{table}{0}
\setcounter{equation}{0}
\renewcommand{\thesection}{S\arabic{section}}
\renewcommand{\thesubsection}{\thesection.\arabic{subsection}}
\renewcommand{\thetable}{S\arabic{table}}
\renewcommand{\thefigure}{S\arabic{figure}}
\renewcommand{\theequation}{S\arabic{equation}}
\def\theHsection{S\arabic{section}}
\def\theHtable{S\arabic{table}}
\def\theHfigure{S\arabic{figure}}
\def\theHequation{S\arabic{equation}}
\titleformat{\section}{\normalfont\Large\bfseries}{\thesection}{1em}{}
\titlespacing*{\section}{0pt}{3.5ex plus 1ex minus .2ex}{2.3ex plus .2ex}
\setlength{\parindent}{0pt}
\setlength{\parskip}{6pt}
\setlength{\tabcolsep}{4pt}
\renewcommand{\arraystretch}{1.10}
\makeatletter
\RenewCommandCopy\caption\arxivSIcaption
\let\@caption\arxiv@SI@caption
\let\@makecaption\arxiv@SI@makecaption
\makeatother
\captionsetup{font=small,labelfont=bf,justification=raggedright,singlelinecheck=false}
\linespread{1}\selectfont

{\Large\bfseries Supplementary Information for Toward an Agentic AI Approach to Public Health Data Collection\par}
\vspace{0.7\baselineskip}

\begingroup\small\tableofcontents\endgroup

\clearpage
\linespread{1}\selectfont
\section{Data collection workflow}\label{sec:technical}

\begin{figure}[p]
\centering
\includegraphics[width=\linewidth,height=0.82\textheight,keepaspectratio]{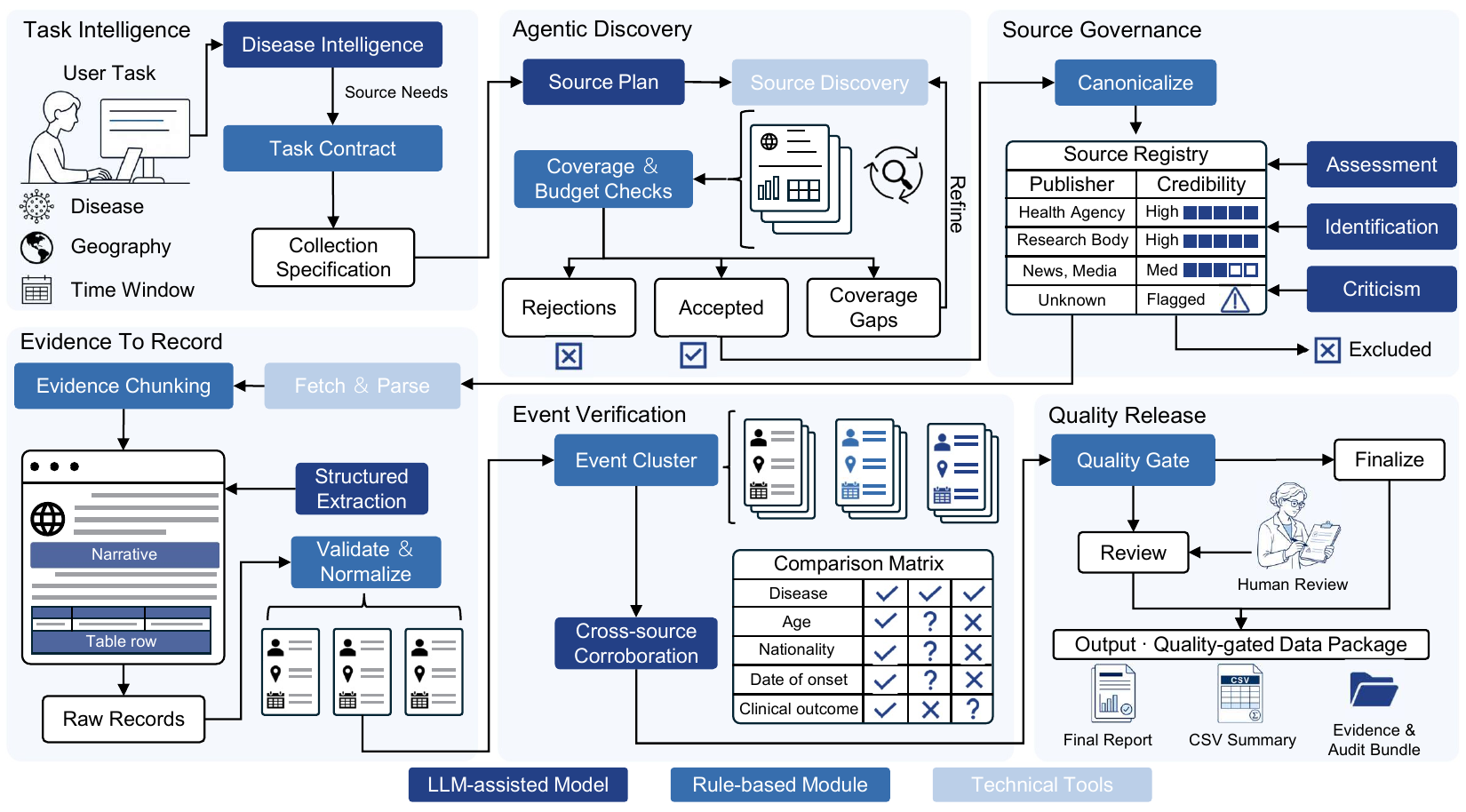}
\caption{\textbf{The general structure of the data collection workflow.} Arrows connect the six stages and show additional search and review paths. Dark blue denotes language model components, medium blue denotes rule based components and light blue denotes retrieval and processing tools. Human-review paths show an optional capability that was not used in the reported studies.}\label{fig:workflow}
\end{figure}

\begin{figure}[p]
\centering
\includegraphics[width=\linewidth,height=0.82\textheight,keepaspectratio]{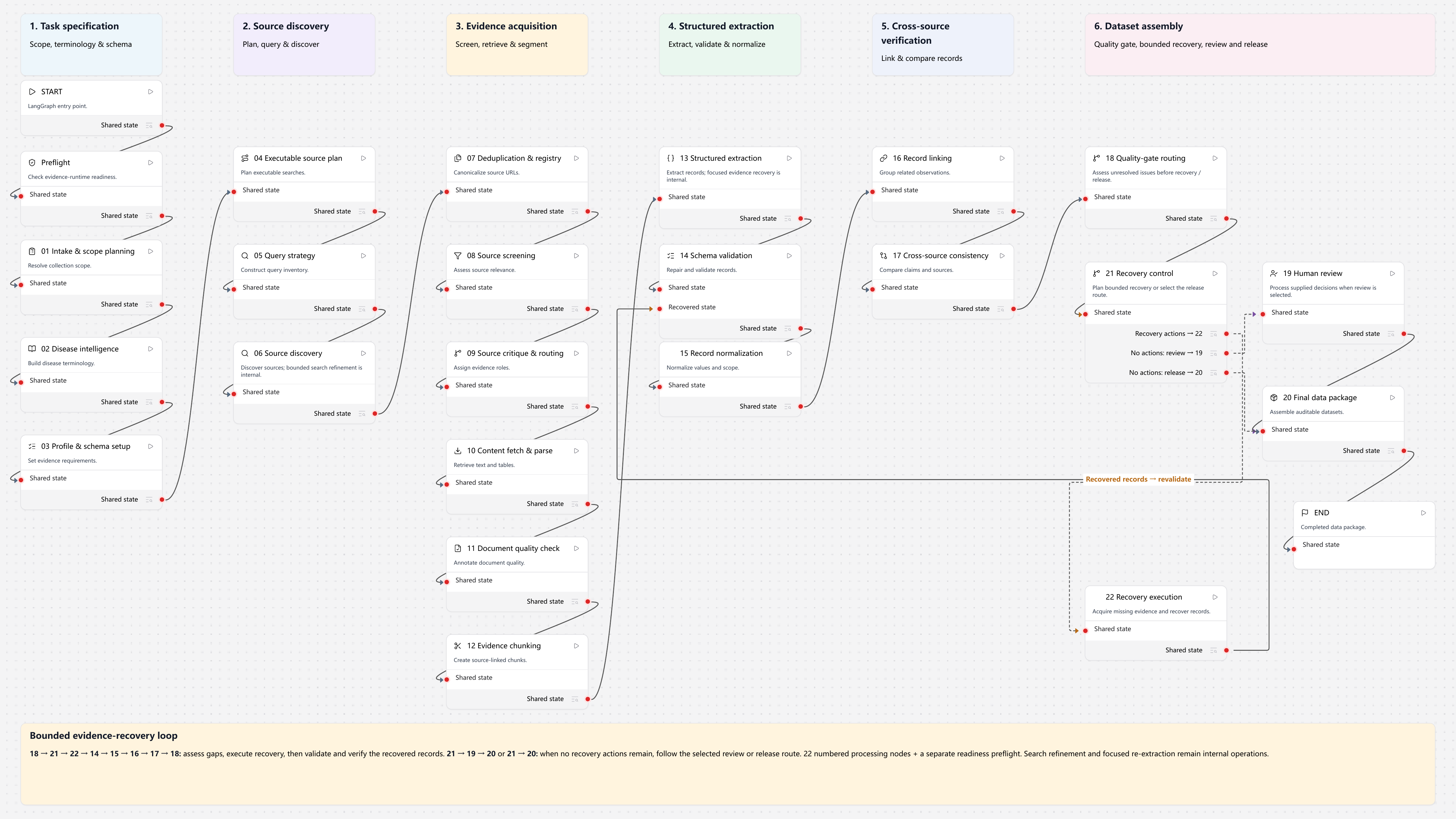}
\caption{\textbf{Langflow visualisation of the workflow.} The canvas groups processing components into six stages. Each node represents a workflow component; its ports indicate the inputs and outputs passed between components. Solid connections show the specified processing sequence, while dashed connections indicate conditional choices. The connections also display the paths for additional evidence collection and optional human review. The human-review capability was not used in the reported studies.}\label{fig:langflow-current}
\end{figure}

The workflow converts public health reports into observations that remain linked to their supporting evidence. It comprises six stages: task specification, source discovery, evidence acquisition, structured extraction, verification across sources, and dataset assembly (Fig.~\ref{fig:workflow}; Table~\ref{tab:nodes}). Language models support interpretation and planning, retrieval tools obtain source material, and deterministic rules govern validation and inclusion. We used the Langflow visualization platform~\cite{supp-langflow} to display the workflow and its components (Fig.~\ref{fig:langflow-current}).

The workflow includes an optional human-review capability. This capability was not used in the reported case studies, and no human review decisions or manual changes to the collected records were applied.

\begingroup\small\setlength{\parskip}{0pt}
\begin{longtable}{@{}>{\raggedright\arraybackslash}p{0.19\linewidth}>{\raggedright\arraybackslash}p{0.47\linewidth}>{\raggedright\arraybackslash}p{0.29\linewidth}@{}}
\caption{Major stages, key activities and outputs of the workflow.}\label{tab:nodes}\\
\toprule
\textbf{Stage} & \textbf{Key activities} & \textbf{Output} \\
\midrule\endfirsthead
\multicolumn{3}{l}{Table \thetable\ (continued)}\\
\toprule
\textbf{Stage} & \textbf{Key activities} & \textbf{Output} \\
\midrule\endhead
\bottomrule\endlastfoot
1. Task specification & Define collection scope; assemble disease terminology; specify requested fields and evidence requirements. & Collection specification and record schema. \\[4pt]
2. Source discovery & Plan source categories and queries; search; refine queries around evidence gaps; consolidate repeated results. & Candidate sources with discovery information. \\[4pt]
3. Evidence acquisition & Assess relevance, publisher and permitted use; retrieve documents; retain extractable text and table context. & Assessed sources and traceable evidence fragments. \\[4pt]
4. Structured extraction & Extract explicitly reported observations; separate statistical meanings; validate and normalise records. & Structured observations with evidence and quality flags. \\[4pt]
5. Verification across sources & Link related reports; identify duplicates; compare aligned observations; retain support and disagreement. & Event groups, supporting evidence and unresolved discrepancies. \\[4pt]
6. Dataset assembly & Assess quality and evidence gaps; perform controlled recovery; retain any supplied review decisions and separate output categories. & Accepted dataset, pending issues and excluded records. \\[4pt]
\end{longtable}\endgroup

\clearpage
\clearpage

\Needspace{10\baselineskip}\subsection{Task specification}

\Needspace{5\baselineskip}

\Needspace{6\baselineskip}
\noindent\textbf{Workflow construction and model configuration.} LangGraph~\cite{supp-langgraph} provides the execution framework that connects the workflow stages, maintains their shared state and selects the next operation. Each node implements a defined processing function, and edges specify either a fixed sequence or a condition for choosing the next function. The shared state carries the task requirements, sources, documents, evidence fragments and record assessments, so later decisions can use the outputs of earlier stages. Figure~\ref{fig:langflow-current} shows the complete workflow. Checkpoints preserve the current state and operation history for an interrupted session; resumption verifies that the task, configuration and processing policies match the saved session.

Users select a supported language model and provide their own API credentials (in this study, we used Claude Sonnet 4.6~\cite{supp-claudesonnet46}). The model adapter passes that configuration to the relevant processing functions and checks compatibility with structured output and generation parameters. Output length, retry limits and sampling settings are configurable; the supplied client settings use a maximum of 4,096 output tokens, at most two provider retries and temperature 0 when supported. These parameters are configurable and do not guarantee identical outputs across models or executions. The settings described below specify execution controls that users can adjust for a collection task.

\noindent\textbf{Collection scope.} Task specification establishes the disease, geography, time window, and requested fields. Explicit constraints guide the interpretation of a free text request. The task contract records accepted disease terms, target geographic level, time bounds and required fields. It separates observations that satisfy the task from broader geographical aggregates, incompatible reporting periods and background material. A disease profile connects the specification to disease names, pathogen and syndrome aliases, exclusion terms, and reporting context. These terms guide search wording and recognition of relevant evidence. Prompt 1 generates this profile and the priorities for source planning from the task. The following boxes present task-specific instructions and user message templates, with excerpts for longer prompts. Ellipses mark omitted content; links identify the complete implementation at a fixed GitHub revision. Double braces identify inputs supplied for a particular task, source or evidence fragment.

\Needspace{7\baselineskip}

\begin{s1promptbox}{Prompt 1: Disease intelligence}
\textbf{System instruction}\par\smallskip

\noindent You create auditable disease intelligence for the data collection workflow. Return only structured source-planning intelligence; do not search the web and do not invent URLs.\par
\medskip\textbf{User message}\par
\begin{lstlisting}[style=s1promptpayload]
{
  "instruction": "Generate disease intelligence for source planning only. Do not perform web search, do not provide URLs, and do not claim source discovery. Return structured terminology, source needs, query terms, validation source categories, extraction priorities, and warnings.",
  "collection_spec": "{{ spec }}",
  "curated_profile_available": "{{ curated_profile is not None }}",
  "curated_profile_hint": "{{ curated_profile or {} }}"
}
\end{lstlisting}
\end{s1promptbox}

\Needspace{10\baselineskip}\subsection{Source discovery}

\Needspace{5\baselineskip}

\noindent\textbf{Source and query planning.} The specification and disease vocabulary are translated into source categories, languages and executable queries. A deterministic plan can be supplemented by model suggestions, after which the proposed queries are checked and consolidated. Query planning combines the task specification, disease vocabulary, selected source categories and search directions. The corresponding prompt keeps proposed queries separate from executed searches and retrieved evidence.

\Needspace{7\baselineskip}

Each planned query specifies its text, search channel, source type, evidence role and rationale, with language and jurisdiction routing fields when needed. A compact response contract retains distinct useful queries while omitting repeated optional metadata. Queries are checked against the task before execution; unsupported disease, geographical or event terms are filtered out.

\begin{s1promptbox}{Prompt 2: Executable source planning}
\textbf{System instruction (excerpt)}\par\smallskip

\noindent You create executable source discovery plans for the data collection workflow. Return only a JSON object matching the ExecutableSourcePlan schema. Do not search the web, fetch URLs, claim that sources were found, or add discovered source URLs.\par
\noindent\ldots\par
\medskip\textbf{User message (excerpt)}\par
\begin{lstlisting}[style=s1promptpayload]
{
  "user_request": "{{ user_request }}",
  "collection_spec": {
    "disease": "{{ spec.get('disease') }}",
    "geography": "{{ spec.get('geography') }}",
    ...
  },
  ...
}
\end{lstlisting}
\smallskip\noindent\textit{Details: see \href{https://github.com/XueyiZhang75/data-collection-workflow/blob/1b619e75914e7df59c521d483a87382665bb00d5/src/data_collection_workflow/nodes/task_scope.py\#L1814-L1902}{GitHub: complete prompt and input template}.}\par
\end{s1promptbox}
\Needspace{5\baselineskip}

\noindent\textbf{Initial search and candidate collection.} Initial planning selects a query batch from the task, source directions and available budget. The workflow then passes the approved query text to Tavily Search~\cite{supp-tavily} through its search adapter. Tavily returns candidate URLs, titles, snippets and available publication metadata, which are recorded with the query and retrieval time. The candidate registry after round $i$ is $\mathcal U_i=\operatorname{Consolidate}\!\left(\mathcal U_{i-1}\cup\bigcup_{q\in\mathcal Q_i}\operatorname{Search}(q)\right)$, where $\mathcal Q_i$ is the executed batch and $\mathcal U_i$ is the retained candidate set for the main iterative search. Consolidation merges repeated reports or canonical URLs while retaining their discovery information. Search snippets guide subsequent collection; the extraction stage uses the retrieved document content. Prompt 3 defines the initial plan that precedes search execution.

\Needspace{7\baselineskip}

The supplied search configuration sets a timeout of 15 seconds, up to eight results per query and four queries per batch. Its fixed allocation permits three main search iterations and up to 20 selected queries; the result cap is the smaller of the general limit of 64 and the iterative limit of 120. Supplementary searches have their own allocations. With adaptive budgeting, the shared session ledger supplies the total limits while batch size remains a scheduling parameter. The next query batch is limited by both the configured batch size and the remaining query allowance. The retained candidate set does not exceed the effective result limit.

\begin{s1promptbox}{Prompt 3: Initial iterative search plan}
\textbf{System instruction (excerpt)}\par\smallskip

\noindent You are planning source discovery for a public-health data collection workflow.\par
\noindent The user task includes disease, location, time window, and target fields. You must decide what to search next based on the task and, for refinement, based on current search-result metadata.\par
\noindent\ldots\par
\medskip\textbf{User message}\par
\begin{lstlisting}[style=s1promptpayload]
{
  "instruction": "Plan the first search iteration. You may reuse/refine upstream planned queries or propose new query text. Do not provide source candidates or page content.",
  "task": "{{ _compact_task(user_request, structured_task, collection_spec) }}",
  "upstream_planned_queries": "{{ planned_queries }}",
  "search_query_inventory": "{{ search_query_inventory }}",
  "bounds": "{{ limits }}"
}
\end{lstlisting}
\smallskip\noindent\textit{Details: see \href{https://github.com/XueyiZhang75/data-collection-workflow/blob/1b619e75914e7df59c521d483a87382665bb00d5/src/data_collection_workflow/agents/iterative_source_discovery_agent.py\#L20-L118}{GitHub: complete prompt and input template}.}\par
\end{s1promptbox}

\Needspace{5\baselineskip}

\Needspace{6\baselineskip}
\noindent\textbf{Refinement around evidence gaps and stopping.} Refinement examines returned result metadata and proposes further queries for remaining gaps. A model stop recommendation is evaluated together with the untried query inventory and available budget. Search stops when an effective search limit is reached, the search provider is disabled or unavailable, no eligible untried query remains, or two consecutive batches yield no new candidates and the required exploration across available source families has been completed.

Eligible queries are untried, task compatible and permitted by the search channel. Effective search limits include the applicable iteration, query and result limits or shared session budget. A model recommendation to stop, or a failed refinement call, does not by itself end this search when eligible queries and budget remain. The runner records the decision and stop reason; bounded exploration does not establish exhaustive coverage.

\Needspace{8\baselineskip}
\noindent\textit{Prompts 3 and 4 use the same complete system instruction; see the GitHub implementation linked in Prompt 3.}\par

\begin{s1promptbox}{Prompt 4: Search refinement and stopping}
\medskip\textbf{User message}\par
\begin{lstlisting}[style=s1promptpayload]
{
  "instruction": "Review the compact search metadata. Decide whether to continue searching or stop. If continuing, return the next bounded query batch. Do not invent facts from snippets.",
  "task": "{{ _compact_task(user_request, structured_task, collection_spec) }}",
  "previous_iteration_plans": "{{ previous_plans }}",
  "all_observations": "{{ observations }}",
  "current_observation": "{{ observation }}",
  "bounds": "{{ limits }}"
}
\end{lstlisting}
\end{s1promptbox}

\Needspace{10\baselineskip}\subsection{Source assessment and evidence acquisition}

\Needspace{5\baselineskip}

\noindent\textbf{Publisher identity and evidence role.} Candidate assessment first distinguishes the organisation responsible for a report from the hosting website and search provider. Publisher, original source and reuse information help identify repeated publication of the same material. Sources may provide collection evidence, supporting evidence or background context, or be reserved for external comparison or excluded. The identity prompt proposes these distinctions from the supplied metadata and any available identity text; deterministic policy determines the permitted use.

\Needspace{7\baselineskip}

\begin{s1promptbox}{Prompt 5: Publisher identity and evidence role}
\textbf{System instruction (excerpt)}\par\smallskip

\noindent You are a source identity assessor for the data collection workflow.\par
\noindent You must not browse, fetch, search, open URLs, or invent facts. Use only the provided source metadata and, when present, already fetched page identity text.\par
\noindent\ldots\par
\medskip\textbf{User message (excerpt)}\par
\begin{lstlisting}[style=s1promptpayload]
{
  "source_entry": "{{ {key: source_entry.get(key) for key in ('source_id', 'canonical_url', 'url', 'domain', 'title', 'snippet', 'publisher', 'source_type', 'search_provider', 'search_result_source_raw', 'search_provider_result_source', 'result_source', 'search_rank', 'query_used', 'discovery_method', 'role_hint', 'source_purpose')} }}",
  "collection_spec": "{{ collection_spec or {} }}",
  "parsed_page_identity": "{{ parsed_page_identity or {} }}",
  ...
}
\end{lstlisting}
\smallskip\noindent\textit{Details: see \href{https://github.com/XueyiZhang75/data-collection-workflow/blob/1b619e75914e7df59c521d483a87382665bb00d5/src/data_collection_workflow/agents/source_identity_agent.py\#L10-L247}{GitHub: complete prompt and input template}.}\par
\end{s1promptbox}

\Needspace{5\baselineskip}

\noindent\textbf{Task relevance and uncertainty.} Screening assesses whether the candidate matches the disease, geography and period and is likely to contain extractable observations. A source critic reviews this assessment against the same task and source use rules. Their confidence scores remain distinct from source credibility. For source $s$, screening combines the two assessments as follows:

\begin{equation}
\begin{aligned}
q_s&=\frac{q_{\mathrm{screen},s}+q_{\mathrm{critic},s}}{2},\\[4pt]
\operatorname{trigger}_{\mathrm{screen}}(s)&=(q_s<0.50)\lor A_s.
\end{aligned}
\label{eq:s1-screening}
\end{equation}

$A_s$ denotes disagreement between the screening and critic decisions. A designation as irrelevant takes precedence and excludes the source; the mean confidence and disagreement rule flags the remaining sources for potential review. Reservation for external comparison and permissions for contextual use are enforced separately. The critic prompt below identifies task mismatches, unsupported extraction uses and uncertainties without making the final policy decision.

\Needspace{7\baselineskip}

Permissions for collection, contextual use and external comparison are determined by the active source policy. The critic receives that policy with the task, so references to reserved authorities in its fixed instruction are interpreted in the context of the specified comparison design.

\begin{s1promptbox}{Prompt 6: Source critic}
\textbf{System instruction (excerpt)}\par\smallskip

\noindent You are a conservative public health source critic agent for the data collection workflow. Assess only the provided source metadata against the active disease, geography, time window, expected public-health data fields, and source role policy. Return advisory JSON only.\par
\noindent\ldots\par
\medskip\textbf{User message (excerpt)}\par
\begin{lstlisting}[style=s1promptpayload]
{
  "source_entry": "{{ source_entry }}",
  "collection_spec": "{{ collection_spec or {} }}",
  "screening_policy": "{{ screening_policy or {} }}",
  ...
}
\end{lstlisting}
\smallskip\noindent\textit{Details: see \href{https://github.com/XueyiZhang75/data-collection-workflow/blob/1b619e75914e7df59c521d483a87382665bb00d5/src/data_collection_workflow/resources/source_critic_agent_prompt.json\#L4-L47}{GitHub: complete task-specific instructions} and \href{https://github.com/XueyiZhang75/data-collection-workflow/blob/1b619e75914e7df59c521d483a87382665bb00d5/src/data_collection_workflow/agents/source_critic_agent.py\#L117-L164}{GitHub: input template and policy handling}.}\par
\end{s1promptbox}

\Needspace{5\baselineskip}

\noindent\textbf{Credibility and advisory review.} A separate weighted rubric summarises nine dimensions: authority, local relevance, disease relevance, timeliness, geographic detail, data detail, readability, independence indicators and provenance. For component scores $x_{sk}$, weights $w_k$ and metadata risk flags $F_s$, the score is

\begin{equation}
\begin{aligned}
p_s&=\min\!\left(0.45,\sum_{\ell=1}^{8}\lambda_\ell\mathbf1\{f_\ell\in F_s\}\right),\\
C_s&=\operatorname{clip}_{[0,1]}\!\left(\sum_{k=1}^{9}w_kx_{sk}-p_s\right).
\end{aligned}
\label{eq:s1-credibility}
\end{equation}

Here $f_\ell$ denotes a named risk and $\lambda_\ell$ its penalty. The weights, in the component order above, are $(0.16,\allowbreak 0.14,\allowbreak 0.16,\allowbreak 0.08,\allowbreak 0.09,\allowbreak 0.14,\allowbreak 0.08,\allowbreak 0.07,\allowbreak 0.08)$. Penalties for ambiguous disease, location and date, unlikely extractability, low authority, secondary news, low readability and screening disagreement are, respectively, $(0.08,\allowbreak 0.06,\allowbreak 0.04,\allowbreak 0.08,\allowbreak 0.06,\allowbreak 0.04,\allowbreak 0.06,\allowbreak 0.10)$. The score is a heuristic assessment, not a probability that the report is correct. The optional advisory prompt below adds role, risk and review suggestions while leaving this deterministic score and source use rules in control.

\Needspace{7\baselineskip}

\begin{s1promptbox}{Prompt 7: Optional credibility advisory}
\textbf{System instruction}\par\smallskip

\noindent You are an advisory source credibility reviewer for the data collection workflow. Return only structured advisory output. Do not browse, fetch, or create URLs.\par
\medskip\textbf{User message}\par
\begin{lstlisting}[style=s1promptpayload]
{
  "instruction": "Review source metadata and deterministic credibility scores. Do not browse, fetch, create URLs, or decide final policy. Return advisory risk flags, role suggestion, and human review recommendation only.",
  "source_entry": "{{ {key: entry.get(key) for key in ('source_id', 'canonical_url', 'title', 'publisher', 'domain', 'source_type', 'discovery_method', 'query_id', 'query_used', 'role_hint', 'snippet')} }}",
  "collection_spec": "{{ state.get('collection_spec') or {} }}",
  "disease_intelligence_summary": "{{ state.get('disease_intelligence_summary') or {} }}",
  "deterministic_assessment": "{{ assessment }}"
}
\end{lstlisting}
\end{s1promptbox}

\Needspace{5\baselineskip}

\Needspace{6\baselineskip}
\noindent\textbf{Retrieval and evidence preparation.} The acquisition module retrieves document content from selected URLs through HTTP requests. It uses browser rendering when a page requires it or when access returns an empty interface or a block; Tavily Extract~\cite{supp-tavilyextract} can provide an alternative when content remains unreadable and external extraction is enabled. Task relevant links and attachments found in retrieved pages can enter the same acquisition process. The parser processes HTML, PDF, tables or JSON according to the response and preserves the source URL, received content, document hash, retrieval time and parser information. Timeouts and response size limits bound each request. PDF pages with fewer than 20 nonspace text characters, or explicitly selected for optical character recognition (OCR), enter OCR at a default render scale of 3. A page is flagged when confidence values are unavailable or any recognised word has confidence below 80. Failed or incomplete retrieval remains available for further collection.

Evidence preparation preserves units within each source: table rows retain their own headers and notes, JSON observations retain an object pointer, and narrative spans retain character offsets and available page or section context. Table text is excluded from narrative splitting to avoid borrowing values across rows. For a narrative span, the target text length is

\begin{equation}
L_e=\max(80,\ 900-h_e),
\label{eq:s1-span-size}
\end{equation}

where $h_e$ is the character length of the bound heading context, including separators. Splitting favours sentence or line boundaries and respects PDF page boundaries when page locators are available. This target applies to narrative preparation, not a uniform table row or structured object length. Field level supporting quotations must resolve to the stored evidence and satisfy the separate 900 character qualification limit.

\Needspace{10\baselineskip}\subsection{Structured extraction and normalisation}

\Needspace{5\baselineskip}

\Needspace{6\baselineskip}
\noindent\textbf{Structured response and extraction budget.} The model adapter selects structured output supported by the provider or a JSON prompt constrained by the extraction schema, then validates the returned object locally with Pydantic~\cite{supp-pydantic}. A malformed response counts as a failed attempt. The supplied scheduler allows four concurrent calls, uses a soft checkpoint at 400 calls and a safety limit of 2,400 calls, and monitors extraction yield over a rolling window of 40 model calls. A tighter session budget takes precedence. At the soft checkpoint, further calls are restricted to evidence with strong record signals or sources that have not met their minimum extraction allocation. The rolling yield is

\begin{equation}
Y=\frac{\sum_{j\in\mathcal W}v_j}{\sum_{j\in\mathcal W}n_j},
\label{eq:s1-extraction-yield}
\end{equation}

Here $\mathcal W$ is the retained rolling window, $v_j$ is the number of valid records returned by an extraction attempt and $n_j$ its actual model calls. Once the window contains 40 calls, yield below 0.05 records per call permits the scheduler to skip a low priority span; spans with strong record signals and sources below their minimum allocation remain eligible. The yield therefore controls further computation without converting a failed extraction into a negative epidemiological observation.

\noindent\textbf{Extraction constrained by evidence.} Rule based parsing, language models or both map the prepared evidence to observation records. The model receives source information, an evidence fragment and the task requirements. Its field level instruction can be summarised as

\begin{equation}
\widehat v_f(e)=
\begin{cases}
v_f(e),&\text{a value for field }f\text{ is explicitly reported in }e,\\
\mathrm{null},&\text{otherwise}.
\end{cases}
\label{eq:s1-evidence}
\end{equation}

$e$ is the supplied evidence, $v_f(e)$ its explicitly stated value for field $f$, and $\widehat v_f(e)$ the requested extraction. This is an instruction to the model rather than a guarantee of correct attribution. A reported zero remains zero; missing values are not inferred. Individual cases, grouped descriptions and aggregate observations are kept distinct, and aggregate counts are not expanded into invented individual records. The extraction prompt specifies these evidence and statistical constraints.

\Needspace{7\baselineskip}

\begin{s1promptbox}{Prompt 8: Evidence-constrained structured extraction}
\textbf{System instruction (excerpt)}\par\smallskip

\noindent You are a public health data extraction agent. Extract only explicitly stated disease case, death, hospitalization, testing, surveillance metric, line-list detail, date, location, and source information from the provided evidence chunk. Use the task only to select relevant evidence; preserve the disease explicitly stated by the source.\par
\noindent\ldots\par
\noindent\textbf{Required output rules (excerpt):}\par
\noindent\ldots\par
\noindent\hangindent=1.2em\hangafter=1 5. Do not infer missing case counts, deaths, dates, countries, or locations.\par
\noindent\hangindent=1.2em\hangafter=1 6. Do not calculate derived quantities such as CFR, duration, or rates unless explicitly stated.\par
\noindent\hangindent=1.2em\hangafter=1 7. Keep confirmed, probable, suspected, and unspecified cases separate.\par
\noindent\ldots\par
\medskip\textbf{User message (excerpt)}\par
\begin{lstlisting}[style=s1promptpayload]
source_id: {{ chunk.get('source_id') }}
source_url: {{ chunk.get('source_url') }}
source_type: {{ chunk.get('source_type') }}
...
Evidence chunk text:
{{ chunk.get('text') or '' }}
\end{lstlisting}
\smallskip\noindent\textit{Details: see \href{https://github.com/XueyiZhang75/data-collection-workflow/blob/1b619e75914e7df59c521d483a87382665bb00d5/src/data_collection_workflow/resources/llm_structured_extraction_policy.json\#L10-L50}{GitHub: base extraction instructions} and \href{https://github.com/XueyiZhang75/data-collection-workflow/blob/1b619e75914e7df59c521d483a87382665bb00d5/src/data_collection_workflow/llm_clients.py\#L691-L770}{GitHub: complete instruction assembly and input template}.}\par
\end{s1promptbox}

\Needspace{5\baselineskip}

\noindent\textbf{Statistical meaning and record context.} Extraction preserves the meaning of a value as well as the value itself. Each observation retains the metric, value, unit, semantic context, reporting period or cutoff, geographic scope, and source and evidence links. Case status and count basis are represented as separate dimensions. Fields may remain missing when the evidence is incomplete. Confirmed, probable, suspected and unspecified cases, as well as cumulative, newly reported, annual and subset counts, remain distinguishable. Publication or report dates, event dates and statistical cutoffs retain their separate meanings. Distinct metrics, periods and scopes produce separate observations when the source supports that distinction.

\Needspace{5\baselineskip}

Each populated factual field is accompanied by a short quotation from the supplied evidence. A table value must belong to the selected row and column, and a patient attribute must belong to that patient passage. The workflow preserves disease, geographical scope and date roles as reported. A populated field passes the evidence check only when the stored text matches its hash, the supporting quotation resolves to the selected evidence fragment and associated structural context, and the quoted value has the appropriate meaning for that field. Document and fragment identifiers, character positions and available table or object locators retain this connection.

\noindent\textbf{Focused evidence recovery.} Evidence spans not represented after the initial extraction enter a focused recovery step. The workflow first applies rule based extraction and, where needed and within the reserved budget, repeats model extraction on the local span. Recovered records remain subject to the same evidence requirements and subsequent validation; this operation remains local to the extraction stage.

\noindent\textbf{Validation and normalisation.} Validation separates unusable records from incomplete records that can be reviewed. Rejection conditions include invalid structure, missing required core information, insufficient minimum content, invalid or unresolved multiple reporting periods, and disease incompatibility. Unresolved review conditions include missing review trigger information or provenance and ambiguous disease relevance. After supported repairs, records are rejected if any rejection condition remains. Otherwise, records with unresolved review conditions are flagged as needing review; the remaining records are classified as valid.

Required core fields are defined by policy and differ from the full set of requested fields; they include disease, source information and a supporting quotation. Provenance also links the record to its evidence fragment, although complete provenance alone does not establish that a value is correctly supported. The test for minimum content does not require a numerical value, and the period checks identify specified inconsistencies rather than every possible ambiguity. Normalization then standardizes supported numeric and date formats while preserving available precision and original evidence. Unsupported values or ambiguous count labels are flagged or cleared, with corrections retained alongside the original observation.

\Needspace{10\baselineskip}\subsection{Verification across sources}

\Needspace{5\baselineskip}

\noindent\textbf{Event linking and duplicate control.} Deterministic linking uses disease, location, temporal and statistical context to define a common event key. Records sharing that key form a candidate group $G$; this groups related reports without resolving patient identities. A candidate group is classified as duplicate only when it contains more than one record, every record has available numeric count content, that content is identical across records, and at least one record provides count semantics information.

The numeric content covers the available confirmed, probable, suspected and unspecified case counts, deaths and hospitalisations. Missing numeric content or absent count semantics leaves related records for further assessment; differing numeric content is retained as a conflict. Incomplete context can still yield a candidate group, so grouping alone does not establish full semantic equivalence. Once the duplicate criteria are met, representative selection follows the ordered preference

\begin{equation}
\begin{aligned}
K_r&=\sum_{f\in\mathcal F_K}\mathbf1\{r_f\ne\mathrm{missing}\}+|\mathcal N(r)|,\\
r^*&\in\operatorname*{arg\,min}_{r\in G}^{\mathrm{lex}}
\bigl(\rho_{\mathrm{role}}(r),\rho_{\mathrm{type}}(r),-C_r,-K_r\bigr).
\end{aligned}
\label{eq:s1-duplicate-preference}
\end{equation}

$\mathcal F_K$ contains disease, country, subnational location, date anchor, report date, source URL, evidence quotation, supporting fragment identifier, statistical count type and count semantics; $\mathcal N(r)$ contains the available numerical count fields. Lower role and source type ranks are preferred, followed by higher credibility $C_r$ and greater completeness $K_r$. The representative enters the deduplicated view, while other records remain linked as supporting evidence. The same count is therefore not contributed again by a duplicate report.

\Needspace{5\baselineskip}

After recovery, linking and consistency checks are repeated on the revised observations, and the current conflict view is rebuilt. Conflict flags represent discrepancies still present in the evidence rather than an accumulation of resolved historical conflicts. The source and recovery history remain available for reassessment.

\noindent\textbf{Discrepancy within an event.} For the available values $x_1,\ldots,x_n$ of the same numerical field in a related event group, absolute and relative spread are

\begin{equation}
\Delta_G=\max_i x_i-\min_i x_i,\qquad
\delta_G=\frac{\Delta_G}{\max(|\min_i x_i|,1)}.
\label{eq:s1-event-difference}
\end{equation}

This comparison requires at least two available values; fewer values do not permit an assessment of consistency. For groups with comparable values, the configured severity rule is

\begin{equation}
\operatorname{severity}(G)=\begin{cases}
\text{equal},&\Delta_G=0,\\
\text{minor},&\Delta_G>0\ \land\ (\Delta_G\leq1\ \lor\ \delta_G\leq0.05),\\
\text{major},&\text{not minor and }\delta_G\geq0.20,\\
\text{intermediate},&\text{otherwise}.
\end{cases}
\label{eq:s1-event-severity}
\end{equation}

Minor differences remain discrepancies in the output; the thresholds organise review priority rather than reconcile the values automatically.

\Needspace{5\baselineskip}

\noindent\textbf{Claim comparison and source support.} Individual reported claims are compared after compatibility checks. For two available numerical values $a$ and $b$, the symmetric relative difference and comparison rule are

\begin{equation}
\begin{aligned}
\delta_{ab}&=\frac{|a-b|}{\max(|a|,|b|,1)},\\
\operatorname{comparison}(a,b)&=\begin{cases}
\text{match},&a=b,\\
\text{minor difference},&|a-b|\leq1\ \lor\ \delta_{ab}\leq0.05,\\
\text{conflict},&\text{otherwise}.
\end{cases}
\end{aligned}
\label{eq:s1-claim-difference}
\end{equation}

The cases are evaluated in order, and missing values yield insufficient evidence. Explicitly incompatible contexts are screened out, although missing temporal or count semantics information does not always prevent comparison. Linked claims form evidence groups that retain corroboration, partial support, duplication from the same source and conflict. Support across sources requires at least two distinct non-empty source identity keys and at least one corroborating or partially supporting claim link.

Source identity draws on an available independence group designation or original publisher, with source identifiers as fallbacks. Conflict takes precedence; background only, exposure only and zero case only groups retain their separate statuses. Among the remaining groups, a corroborating pair distinguishes corroboration from partial support. Distinct identities describe provenance rather than proven statistical independence. External reference comparisons remain separate from corroboration among collected sources.

\Needspace{10\baselineskip}\subsection{Dataset assembly}

\Needspace{5\baselineskip}

\Needspace{6\baselineskip}
\noindent\textbf{Quality gating and inclusion.} The workflow qualifies observations against evidence within their original source before release. For each populated factual field, it verifies the document and evidence locator, the quoted value and its semantic role. It also checks task scope, requires supported geography, and requires an observation period or date field for numeric observations. Unmet evidence requirements include unsupported fields, missing required geographical or temporal information, and violations of task constraints. A record is classified as qualified when no unmet evidence requirements remain; otherwise, it is retained as a candidate.

A supported patient identifier distinguishes an individual case record; a numerical observation without such an identifier is classified as aggregate. Other supported material is retained as context. Candidate records preserve unresolved reasons and enter recovery or pending review outputs. Qualification is a rule based evidence assessment, not expert approval; unsupported fields cannot be justified by the task text, search query or publisher metadata.

\noindent\textbf{Controlled evidence recovery.} After quality assessment, the workflow checks unmet source and field requirements, incomplete acquisition and conflicting evidence. Current evidence, remaining budget and action history determine which recovery actions can be executed for the current workflow state. If executable actions remain, the controller performs recovery. Otherwise, it selects the optional human-review route when human review is enabled and the configured review condition is met, or proceeds to dataset assembly.

Recovery performs the planned source assessment, search, retrieval, parsing or extraction, then merges the resulting evidence. Recovered observations return to validation, normalization, linking and consistency checks before quality is reassessed. Available actions, budgets and progress checks bound this cycle; unresolved issues remain in the output.

The controller plans at most eight recovery actions per pass from remaining gaps, budget and action history. Completed actions are not repeated, and a failed action can receive at most two actual attempts. Strict mode allows at most two recovery rounds; adaptive mode instead uses the shared budget and progress conditions. Every pass recomputes evidence qualification and coverage, and records whether recovery stops because requirements are met, budgets are exhausted, the provider is restricted or no executable action remains.

\Needspace{6\baselineskip}
\noindent\textbf{Optional human review across the workflow.} The workflow supports optional human review of uncertain sources, ambiguous records and conflicting evidence. Source assessment, schema validation, normalization, and cross-source verification generate review flags, which are consolidated with supporting evidence for potential expert assessment. When enabled, reviewers can accept, reject, or revise records, or request additional evidence, with decisions recorded and affected records reassessed. Human review was disabled in all three case studies, and no human review decisions or manual record modifications were applied during data collection.

\clearpage 
\linespread{1}\selectfont
\section{Case study of hantavirus in 2026}\label{sec:hantavirus}

\subsection{Case overview and evaluation design}\label{sec:hantavirus-overview}

We evaluated source coverage and recovery of information for each case for the 2026 MV Hondius-associated hantavirus outbreak. The study window was 1 April--30 June 2026. The evaluation cohort comprised 13 numbered cases: 12 confirmed and one probable. The final workflow case study dataset and the manually curated Global.health/Hondius GitHub benchmark~\cite{supp-globalhealth2026} both represented all 13 cases. All coverage and field quality results below refer to this cohort. WHO release counts shown in main-text Fig.~2a were compiled from ten WHO publications released between 4 and 28 May 2026~\cite{supp-whodon599,supp-whonote0507,supp-whodon600,supp-whodg0512,supp-whodon601,supp-whodg0515,supp-si-whorra,supp-whodg0519,supp-whodg0522,supp-whohantavirus2026}. We assessed source coverage using distinct URLs, source domains and associations between cases and URLs, and evaluated field quality through a paired comparison of the fields for each case. Both datasets were scored against the same curated clinical reference. This reference combined the published case table~\cite{supp-si-eurosurveillance}, its 9 July 2026 addendum~\cite{supp-si-eurosurveillance-addendum} and WHO risk-assessment material~\cite{supp-si-whorra}.

\begin{center}\makebox[\linewidth][c]{\includegraphics[width=160mm]{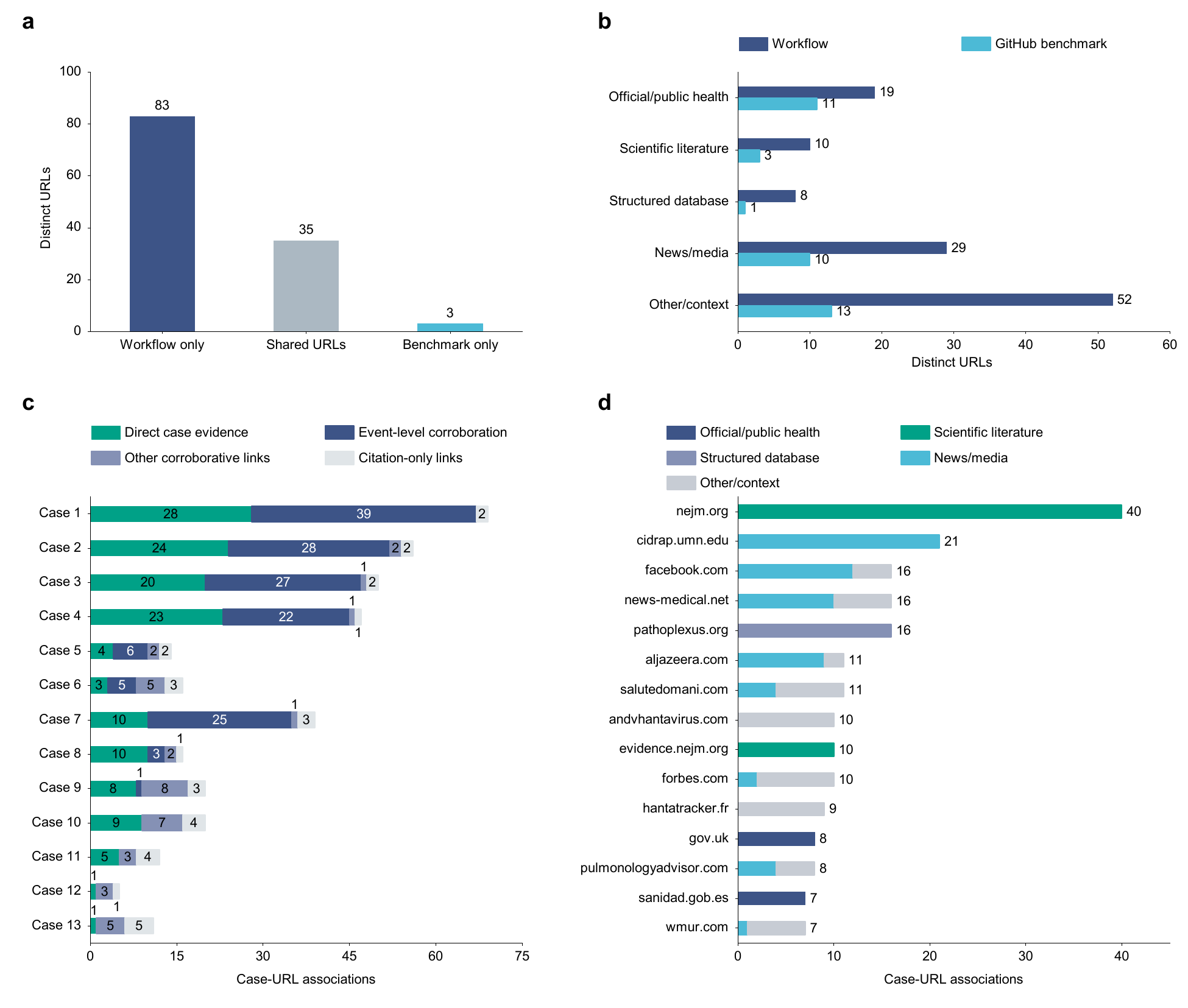}}\end{center}
\captionof{figure}{\textbf{Source coverage and evidence scope for the hantavirus case study.} a, URLs found only in the workflow, in both datasets, or only in the GitHub benchmark. b, Counts by source category. In a--b, both datasets use the same URL normalization rule, and each URL is counted once per dataset. c, Associations between cases and URLs by evidence type. Direct evidence supports a specific case; event level evidence links consistent case details within the same outbreak. Other links connect patients, groups, sequences or reports from different sources. d, The 15 hostnames with the most associations, coloured by source category. In c--d, a URL linked to several cases is counted once for each case.}\label{fig:s3}

\loadgeometry{si}
\clearpage

\Needspace{5\baselineskip}\subsection{Source coverage and evidence scope}

The final source inventory contained 375 associations between cases and URLs across 85 source domains. A URL associated with several cases contributed one association per case. For comparison with the GitHub benchmark, we applied the same URL normalization rule to both datasets. Hostnames were lowercased; an initial www., the trailing path slash and fragment identifiers were removed. Query strings and HTTP/HTTPS distinctions were retained. The 132 URL strings in the workflow inventory corresponded to 118 distinct URLs under this rule. The benchmark contained 38 distinct URLs; 35 were shared, 83 occurred only in the workflow dataset and three only in the benchmark (Fig.~\ref{fig:s3}a).

Workflow sources covered all five source classes (Fig.~\ref{fig:s3}b). The workflow dataset contained 19 official or public health URLs and ten URLs from scientific literature, compared with 11 and three in the benchmark. Structured databases, news reports and other contextual sources supplied the remaining URLs. These categories describe the role of each webpage. Each normalized URL was counted once per dataset and assigned a single class. URLs with both media and contextual labels were assigned to the more specific news/media class.

The evidence associated with individual cases varied in scope (Fig.~\ref{fig:s3}c). Direct case evidence comprised 146 associations (38.9\%) with excerpts supporting a specific numbered case. Event level corroboration comprised 156 associations (41.6\%) linked through compatible case attributes within the same outbreak. Another 40 associations (10.7\%) involved patient profiles, case groups, sequence accessions or other corroborative links. Direct evidence covered all 13 cases, with one to 28 associations per case; cases 12 and 13 each had one direct association. Restricting the inventory to direct evidence retained all 13 cases across 39 domains, with a median of seven domains per case, compared with 14 in the full inventory.

The 15 most frequently represented hostnames accounted for 200 of the 375 associations (53.3\%); the remaining 70 accounted for 175 (46.7\%; Fig.~\ref{fig:s3}d). The largest contributors were nejm.org (40 associations) and cidrap.umn.edu (21), followed by facebook.com, news-medical.net and pathoplexus.org (16 each). These counts describe how reports linked to cases are distributed across websites.

\clearpage
The broader source inventory was also visible within individual cases (Table~\ref{tab:hantavirus-breadth}). The workflow linked more distinct domains than the benchmark for all 13 cases; the medians were 14 and four domains per case, respectively. After restricting workflow links to direct case evidence, the median remained seven domains and the domain count exceeded the benchmark for eight cases.

\begingroup\small\setlength{\tabcolsep}{4pt}\renewcommand{\arraystretch}{1.15}
\begin{longtable}{@{}>{\raggedright\arraybackslash}p{27mm}rrr@{}}
\caption{Source domains linked to each case in the hantavirus evaluation cohort.}\label{tab:hantavirus-breadth}\\
\toprule
Case & \shortstack{Workflow\\all domains} & \shortstack{Workflow\\direct evidence} & \shortstack{Benchmark\\all domains} \\
\midrule\endfirsthead
\toprule
Case & \shortstack{Workflow\\all domains} & \shortstack{Workflow\\direct evidence} & \shortstack{Benchmark\\all domains} \\
\midrule\endhead
Case 1 & 54 & 19 & 3 \\
Case 2 & 41 & 14 & 4 \\
Case 3 & 35 & 12 & 4 \\
Case 4 & 32 & 14 & 2 \\
Case 5 & 12 & 3 & 4 \\
Case 6 & 11 & 2 & 5 \\
Case 7 & 32 & 8 & 4 \\
Case 8 & 12 & 7 & 4 \\
Case 9 & 14 & 7 & 5 \\
Case 10 & 14 & 7 & 6 \\
Case 11 & 9 & 4 & 5 \\
Case 12 & 5 & 1 & 3 \\
Case 13 & 9 & 1 & 6 \\
Median & 14 & 7 & 4 \\
\bottomrule
\end{longtable}
\endgroup

The workflow therefore provides a broader collection of source material for reconstructing the outbreak. Its 118 distinct URLs span official reports, scientific literature, news and other sources, and its direct evidence covers all 13 cases. The following analysis evaluates the recovery of individual fields from this material.

\clearpage

\Needspace{5\baselineskip}\subsection{Recovery of case information}

We evaluated nine attributes for 13 cases (117 entries per dataset) against the same clinical reference: the published case table, its 9 July addendum and WHO risk assessments. A populated value agreeing with the reference is correct, a conflicting value incorrect, and an absent assessable value missing. Values without a comparable reference or not applicable to the case are not evaluable. These comprise all location and exact-age values, sex for three cases and confirmation date for one: 30 entries excluded from the common denominator of 87 (Fig.~\ref{fig:s4}a,b).

The workflow contained 59 correct, 11 incorrect and 17 missing values; the benchmark contained 74, ten and three, respectively. Table~\ref{tab:hantavirus} reports correct recovery, completeness and correctness among populated fields.

\par\Needspace{.32\textheight}\begingroup\small\setlength{\parskip}{0pt}\renewcommand{\arraystretch}{1.15}
\begin{longtable}{@{}>{\raggedright\arraybackslash}p{.39\linewidth}>{\raggedright\arraybackslash}p{.285\linewidth}>{\raggedright\arraybackslash}p{.285\linewidth}@{}}
\caption{Overall field quality for the hantavirus case study datasets.}\label{tab:hantavirus}\\
\toprule
\textbf{Measure} & \textbf{Workflow} & \textbf{GitHub benchmark} \\
\midrule\endfirsthead
\toprule
\textbf{Measure} & \textbf{Workflow} & \textbf{GitHub benchmark} \\
\midrule\endhead
\bottomrule\endlastfoot
Correct recovery & 59/87 (67.82\%) & 74/87 (85.06\%) \\[5pt]
Field completeness & 70/87 (80.46\%) & 84/87 (96.55\%) \\[5pt]
Correctness among populated fields & 59/70 (84.29\%) & 74/84 (88.10\%) \\[5pt]
\end{longtable}
Correct recovery = correct/evaluable; completeness = populated/evaluable; correctness among populated fields = correct/populated. All denominators exclude the 30 non-evaluable cells.
\endgroup

Correct recovery varied by field (Fig.~\ref{fig:s4}c). Case status was correct for all 13 cases in both datasets. The workflow correctly recovered nationality for 12/13 cases, outcome for 11/13, sex for 8/10 and confirmation date for 7/12. Age group and onset date each had four correct values among 13 evaluable cases. These two fields accounted for 13 of the 17 missing values, while onset and confirmation dates accounted for seven of the 11 incorrect values. The field level differences therefore centred on missing age-group information and incomplete or inconsistent clinical dates.

The paired comparison identified 56 cells that both datasets recovered correctly, 18 correct only in the benchmark, three correct only in the workflow dataset and ten correct in neither. The benchmark therefore recovered 15 additional values overall. Age group and onset date accounted for 11 of this 15-value difference.

\clearpage\newgeometry{left=12mm,right=12mm,top=18mm,bottom=18mm,headheight=14pt}

\begin{center}\makebox[\linewidth][c]{\includegraphics[width=180mm]{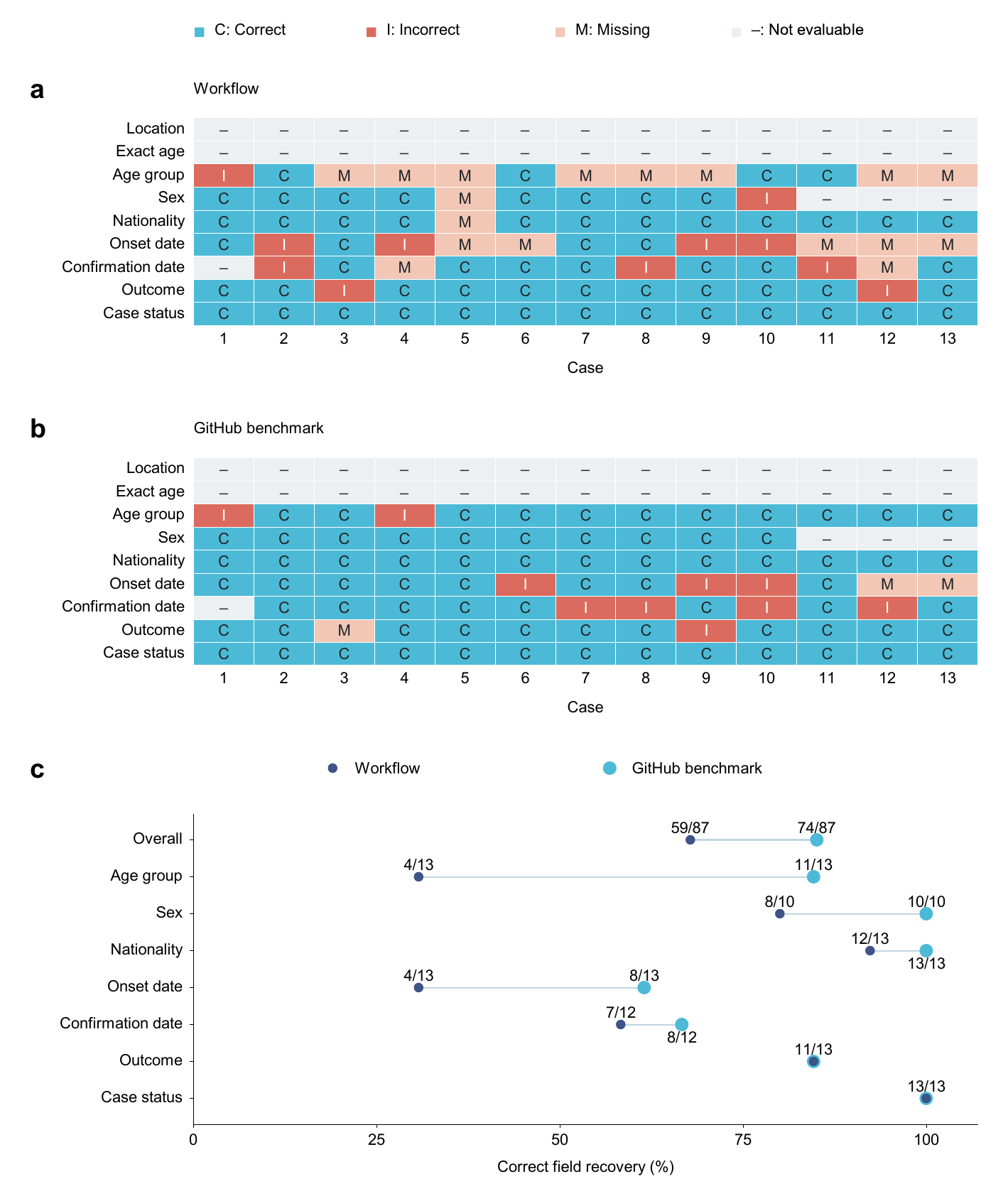}}\end{center}
\captionof{figure}{\textbf{Field level results for the hantavirus case study.} a,b, Scores for the workflow dataset and GitHub benchmark across 13 cases and nine fields. C, correct; I, incorrect; M, missing; dash, not evaluable. Each dataset contains 87 evaluable and 30 non-evaluable cells. c, Correct values as a percentage of evaluable cells, overall and by field. Labels show correct/evaluable counts, and lines connect the two datasets for each field. Equal values share one position and one label. The overall denominator is 87; field denominators include only cells with an evaluable reference value.}\label{fig:s4}

\loadgeometry{si}

\clearpage
\linespread{1}\selectfont
\section{Case study of measles in Canada in 2025}\label{sec:canada}

\Needspace{13\baselineskip}\subsection{Case overview and comparison design}

\begin{samepage}
We evaluated workflow source coverage, cumulative count agreement and reporting times at national and provincial scales using Canadian public measles reports from 2025. For this case study, we combined provincial, territorial and geographically attributable local reports. Within each jurisdiction, overlapping cumulative records contributed the highest count available by the comparison date. Counts for explicitly non-overlapping components were summed. The larger of the jurisdiction-wide count and the sum of these components contributed to the national total. Values were carried forward between updates, with dates before the first eligible observation treated as missing. PHAC reports and reports that explicitly attributed their national counts to PHAC were excluded from the workflow series and retained only as reference material.

\end{samepage}

\begin{samepage}
The national reference comprised 50 reference comparison dates from the Public Health Agency of Canada (PHAC)~\cite{supp-phac_dec2025} and institutional reports explicitly attributing national totals to PHAC~\cite{supp-nysdoh_ghu,supp-paho_epiupdate,supp-paho_bulletin}. PHAC-derived national values were assigned to the reference series. Workflow observations retained their reported confirmed, probable, combined or unspecified case definitions. The reference used combined confirmed-plus-probable counts where available. Its first six reporting weeks contained confirmed counts only, and its reporting year included the week beginning 29 December 2024. These comparisons measure agreement between public-reporting series with the case definitions and reporting periods specified above. Timing analyses used publication dates, with date limits recorded when the exact date was uncertain. Statistical cutoff dates identified the periods represented by the counts.

\end{samepage}

\Needspace{12\baselineskip}\subsection{Source composition and geographical coverage}

\begin{samepage}
The evidence set contained 217 source observations, of which 182 (83.9\%) came from official sources and 35 (16.1\%) from attributed secondary reports. Official observations formed the basis of the workflow series, supplemented by news outlets and other reports with identifiable attribution. The inventory contained 33 publisher or institutional source labels: 20 official and 13 secondary. These labels retained departmental and joint-publisher attributions as recorded; each observation was counted once.

\end{samepage}

\begin{samepage}
Observations covered all 13 provinces and territories, with different numbers of observations across jurisdictions (Fig.~\ref{fig:canada-sources}). Alberta, British Columbia, Ontario and Manitoba contributed 53, 48, 42 and 31 observations, respectively, together accounting for 174/217 records (80.2\%). Major contributors included the Alberta government (43 observations), the BC Centre for Disease Control (39), Public Health Ontario (39) and Manitoba Health (24). Secondary publishers supplied information for several jurisdictions, supplementing the provincial sources shown in the figure.

\end{samepage}

\par\medskip\noindent\begin{minipage}{\linewidth}
\begin{center}\makebox[\linewidth][c]{\includegraphics[width=190mm]{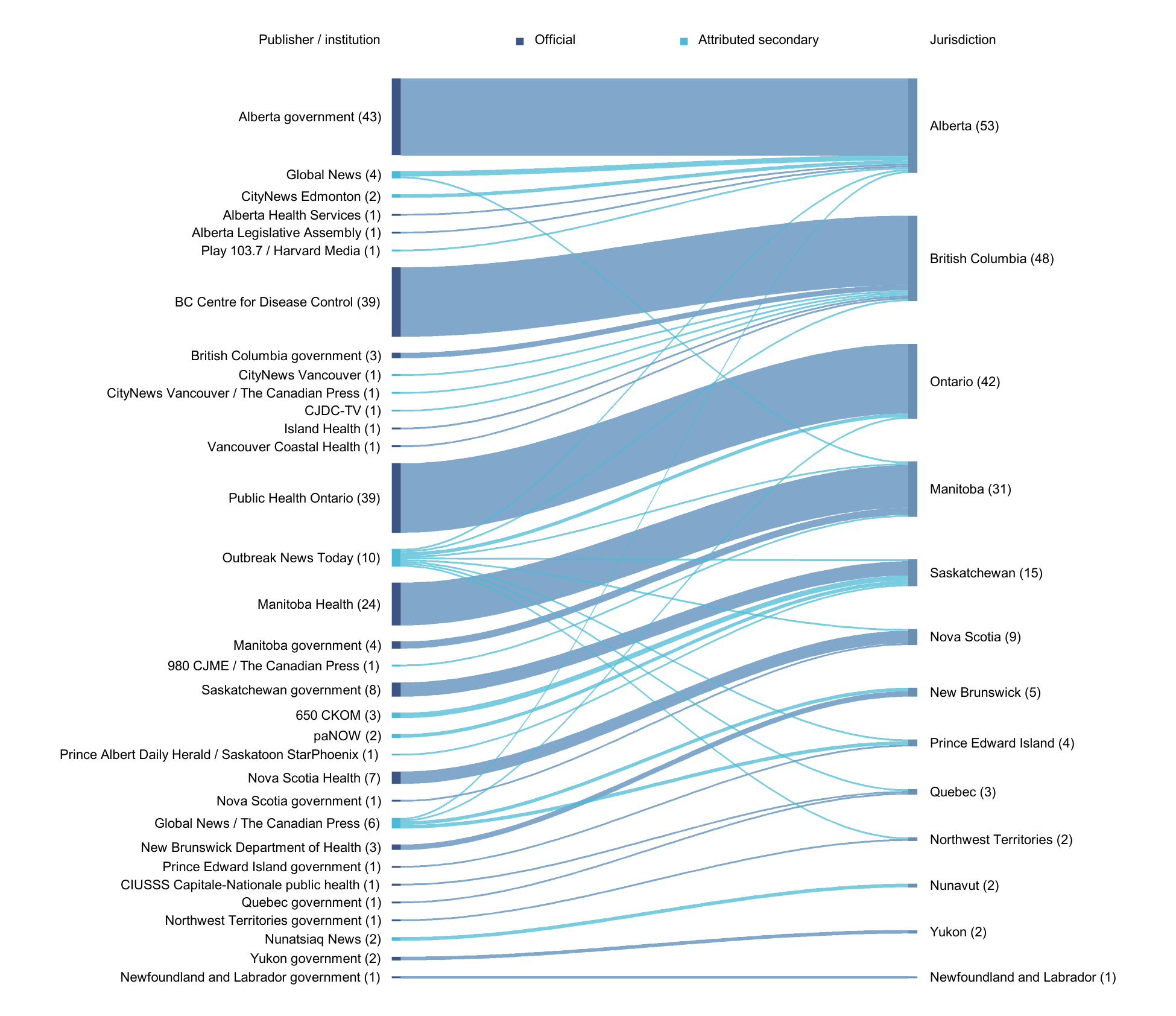}}\end{center}
\captionof{figure}{\textbf{Source coverage across Canadian provinces and territories.} The bands link 33 source labels on the left to 13 provinces and territories on the right. Band widths show the number of observations, with 217 observations in total. Blue represents official sources (182 observations); cyan represents secondary reports with named original sources (35). Numbers beside each source or jurisdiction show its total. Source names retain the recorded departmental or joint publisher and news-agency credits. Each observation is counted once, including reports of zero cases.}\label{fig:canada-sources}
\end{minipage}\par\medskip

\clearpage
\begin{samepage}
The monthly distribution showed differences in record counts and source composition (Table~\ref{tab:canada-monthly-source}). June and July contained 32 and 34 records, respectively. Secondary reports supplied 17 of the June observations, whereas 31 of the 34 July observations were official. All 76 observations assigned to August--November were official. Records were assigned to months using the dates on which the information was known to be available, including date limits supported by archives or webpage updates. The dataset contained no eligible January observation.

\end{samepage}

\par\medskip\par\medskip

\par\medskip\begingroup\small\renewcommand{\arraystretch}{1.12}
\begin{longtable}{@{}lrrr@{}}
\caption{Canadian measles: monthly source observations.}\label{tab:canada-monthly-source}\\
\toprule
Month & Official & Secondary & Total \\
\midrule\endfirsthead
\toprule
Month & Official & Secondary & Total \\
\midrule\endhead
Jan & 0 & 0 & 0 \\
Feb & 6 & 2 & 8 \\
Mar & 9 & 4 & 13 \\
Apr & 11 & 3 & 14 \\
May & 20 & 5 & 25 \\
Jun & 15 & 17 & 32 \\
Jul & 31 & 3 & 34 \\
Aug & 20 & 0 & 20 \\
Sep & 22 & 0 & 22 \\
Oct & 21 & 0 & 21 \\
Nov & 13 & 0 & 13 \\
Dec & 14 & 1 & 15 \\
\midrule
Total & 182 & 35 & 217 \\
\bottomrule\end{longtable}\endgroup

Each retained observation is counted once. Months use recorded public availability dates, including documented upper bounds when necessary.

\clearpage
\subsection{Duration and magnitude of reported count}\label{sec:canada-count-advantage}

To characterize the duration and magnitude of differences between the workflow and reference reporting series, we compare the two retained reporting series on every calendar day. Let $C_P$ contain the days in period $P$ on which both counts are available. We define the signed difference, number and proportion of days with higher workflow counts, and mean daily difference as:

\begin{equation}
D_t=W_t-R_t,\quad A_P=\sum_{t\in C_P}\mathbf{1}(D_t>0),\quad
p_P^{\rm day}=100\,\frac{A_P}{|C_P|},\quad
\overline{D}_P^{\rm day}=\frac{1}{|C_P|}\sum_{t\in C_P}D_t.
\label{eq:reported-count-lead}
\end{equation}

The relative count difference scales the signed mean by the reference mean over the same comparable dates:

\begin{equation}
\rho_P=100\,\frac{\sum_{t\in C_P}(W_t-R_t)}{\sum_{t\in C_P}R_t}
=100\,\frac{\overline{D}_P^{\rm day}}{\overline{R}_P^{\rm day}}.
\label{eq:relative-count-difference}
\end{equation}

Here $\overline{R}_P^{\rm day}$ is the mean reference count over $C_P$. Both count difference measures include positive, zero and negative differences. Relative $\rho_P$ is not the proportion of higher-count days or an unweighted mean of daily percentages. It is undefined when no comparable dates exist or the reference denominator is zero.

The daily states follow the same source selection and aggregation rules as the national comparison. A report enters on its publication date or, where only a date bound is supported, the documented date by which it was available. Each state is held forward until a later eligible update; future observations are not interpolated or backfilled. Dates before both series have an eligible value remain unavailable. These comparisons represent retrospectively reconstructed public reporting, not the actual acquisition times of an operational surveillance system. Moreover, the two series may differ in reporting frequency, statistical cutoffs, and case definitions. Consequently, higher workflow counts do not necessarily indicate earlier case detection, additional unique cases or greater epidemiological accuracy.

Canada had 331 comparable days, beginning on 4 February. Workflow counts were higher on 264/331 days (79.8\%), with a mean signed daily difference of +88.71 cases, corresponding to a relative difference of +2.77\%. In April the count was higher on all 30 days; in May it was higher on 29/31 days, with a mean difference of +326.32 cases. The first three and six tracking months had 69/89 (77.5\%) and 157/181 (86.7\%) higher-count days, respectively, and mean daily differences of +80.29 and +143.63 cases. The first nine months had 225/273 higher-count days (82.4\%), a mean difference of +105.82 cases and a relative difference of +3.83\%. December retained 14/31 higher-count days and a negative mean of -3.81 cases. These results indicate that the reconstructed workflow series generally contained higher reported cumulative counts, particularly during the earlier months, but that the magnitude and persistence of these differences diminished or varied later in the year.

The 34 calendar-year days before the common comparison starts remain outside the comparable-day denominator. Relative to all 365 calendar days, documented higher-count days occupy 72.3\% of the year; the remaining days include both lower-count and unavailable periods. This coverage distinction is retained in the tables.

Figure~\ref{fig:canada-window-metrics}a,b pairs higher-count duration with its fraction of comparable days; c,d pairs the mean signed daily count difference with its relative percentage. Monthly and tracking-window results are shown separately. All panels include the same available-by observations without filtering for exact threshold dates. Tables~\ref{tab:canada-leading-case-monthly} and \ref{tab:canada-leading-case-windows} separately report duration, coverage and exact values. Because the series are carried forward between reports, consecutive daily comparisons and overlapping tracking windows are temporally dependent and should be interpreted descriptively rather than as independent evidence of improved surveillance performance.

\clearpage\newgeometry{left=15mm,right=15mm,top=18mm,bottom=18mm,headheight=14pt}
\begin{center}\includegraphics[width=180mm]{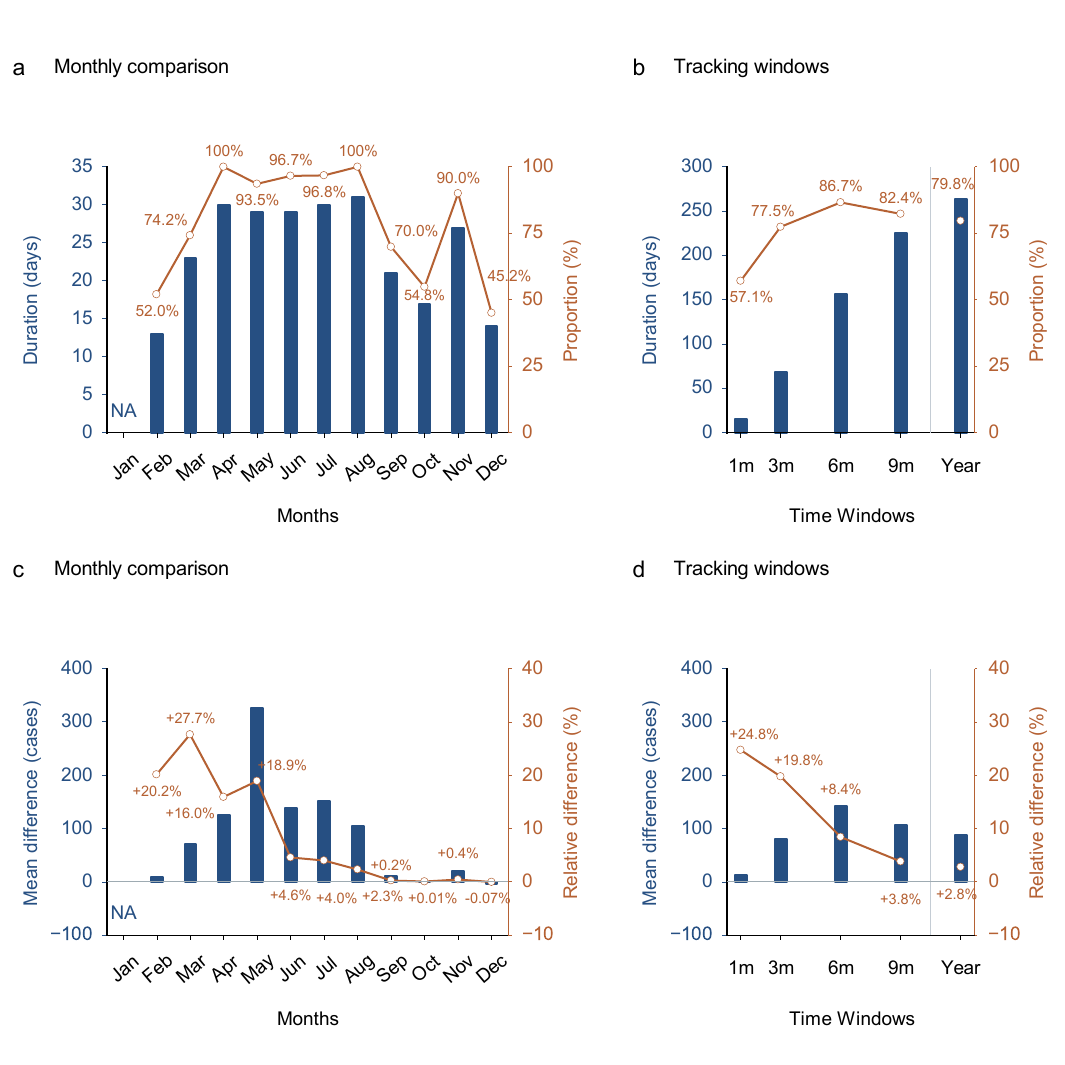}\end{center}
\captionof{figure}{\textbf{Duration and magnitude of reported-count advantage for Canadian measles.} a,b, Days with $W>R$ (blue bars, left axis) and their proportion among comparable days (orange line, right axis). c,d, Mean signed $W-R$ (blue bars, left axis) and relative difference $\rho$ (orange line, right axis; Eq.~\ref{eq:relative-count-difference}). a,c, Calendar months. b,d, First one, three, six and nine tracking months, with Year shown separately. Only orange percentages are labelled. Positive differences indicate higher workflow-reported counts. NA denotes no comparable days; zero is an observed value. Duration is not mean reporting-date advance.}\label{fig:canada-window-metrics}
\loadgeometry{si}\clearpage

\begingroup\small\setlength{\tabcolsep}{4pt}\renewcommand{\arraystretch}{1.15}
\begin{longtable}{@{}>{\raggedright\arraybackslash}p{27mm}>{\centering\arraybackslash}p{26mm}>{\centering\arraybackslash}p{30mm}>{\centering\arraybackslash}p{18mm}>{\centering\arraybackslash}p{22mm}>{\centering\arraybackslash}p{23mm}@{}}
\caption{Canadian measles: daily reported-count advantage by calendar period.}\label{tab:canada-leading-case-monthly}\\
\toprule
Period & \shortstack{Comparable /\\calendar days} & \shortstack{Higher / equal\\/ lower days} & \shortstack{Higher\\(\%)} & \shortstack{Mean $D$\\(cases)} & \shortstack{Relative $\rho$\\(\%)} \\
\midrule\endfirsthead
\toprule
Period & \shortstack{Comparable /\\calendar days} & \shortstack{Higher / equal\\/ lower days} & \shortstack{Higher\\(\%)} & \shortstack{Mean $D$\\(cases)} & \shortstack{Relative $\rho$\\(\%)} \\
\midrule\endhead
Jan & 0/31 & --- & --- & --- & --- \\
Feb & 25/28 & 13/0/12 & 52.0 & +8.80 & +20.16 \\
Mar & 31/31 & 23/0/8 & 74.2 & +70.65 & +27.68 \\
Apr & 30/30 & 30/0/0 & 100.0 & +125.90 & +15.98 \\
May & 31/31 & 29/0/2 & 93.5 & +326.32 & +18.94 \\
Jun & 30/30 & 29/0/1 & 96.7 & +139.57 & +4.55 \\
Jul & 31/31 & 30/0/1 & 96.8 & +151.81 & +3.99 \\
Aug & 31/31 & 31/0/0 & 100.0 & +105.03 & +2.32 \\
Sep & 30/30 & 21/0/9 & 70.0 & +12.20 & +0.25 \\
Oct & 31/31 & 17/0/14 & 54.8 & +0.71 & +0.01 \\
Nov & 30/30 & 27/0/3 & 90.0 & +21.40 & +0.41 \\
Dec & 31/31 & 14/0/17 & 45.2 & -3.81 & -0.07 \\
Year & 331/365 & 264/0/67 & 79.8 & +88.71 & +2.77 \\
\bottomrule
\end{longtable}
\endgroup

The higher-day percentage divides by comparable days; relative $\rho$ divides the mean signed difference by the reference mean over the same dates (Eq.~\ref{eq:relative-count-difference}). Both count difference measures retain positive, zero and negative daily differences. A dash denotes no eligible comparison or a zero reference denominator; observed zero is reported as zero. Year retains 365 days in the coverage column.

\clearpage
\begingroup\small\setlength{\tabcolsep}{4pt}\renewcommand{\arraystretch}{1.15}
\begin{longtable}{@{}>{\raggedright\arraybackslash}p{27mm}>{\centering\arraybackslash}p{26mm}>{\centering\arraybackslash}p{30mm}>{\centering\arraybackslash}p{18mm}>{\centering\arraybackslash}p{22mm}>{\centering\arraybackslash}p{23mm}@{}}
\caption{Canadian measles: daily reported-count advantage by tracking window.}\label{tab:canada-leading-case-windows}\\
\toprule
Period & \shortstack{Comparable /\\calendar days} & \shortstack{Higher / equal\\/ lower days} & \shortstack{Higher\\(\%)} & \shortstack{Mean $D$\\(cases)} & \shortstack{Relative $\rho$\\(\%)} \\
\midrule\endfirsthead
\toprule
Period & \shortstack{Comparable /\\calendar days} & \shortstack{Higher / equal\\/ lower days} & \shortstack{Higher\\(\%)} & \shortstack{Mean $D$\\(cases)} & \shortstack{Relative $\rho$\\(\%)} \\
\midrule\endhead
First month & 28/28 & 16/0/12 & 57.1 & +12.46 & +24.77 \\
First 3 months & 89/89 & 69/0/20 & 77.5 & +80.29 & +19.82 \\
First 6 months & 181/181 & 157/0/24 & 86.7 & +143.63 & +8.43 \\
First 9 months & 273/273 & 225/0/48 & 82.4 & +105.82 & +3.83 \\
Year & 331/365 & 264/0/67 & 79.8 & +88.71 & +2.77 \\
\bottomrule
\end{longtable}
\endgroup

The higher-day percentage divides by comparable days; relative $\rho$ divides the mean signed difference by the reference mean over the same dates (Eq.~\ref{eq:relative-count-difference}). Both count difference measures retain positive, zero and negative daily differences. A dash denotes no eligible comparison or a zero reference denominator; observed zero is reported as zero. Year retains 365 days in the coverage column.

Tracking windows start on 4 February 2025. The first one, three, six and nine months end on 3 March, 3 May, 3 August and 3 November, respectively. Year covers 1 January--31 December. The nested tracking windows overlap.

\Needspace{9\baselineskip}
\noindent\textbf{Reporting-time context.} For a count $q$, $T_X(q)$ is the date of the first eligible report in series $X$ reaching at least $q$; $T_R(q)-T_W(q)>0$ denotes earlier workflow reporting. Bounded dates establish order only when the bounds suffice; exact means use no substituted dates. Across 50 reference levels, 22 were earlier, 15 later and two the same day; ten were undetermined and one unreached. The 28 exact pairs averaged $-2.29$ days; 11 of the 22 earlier levels lacked exact pairs. The workflow recorded 881 cases on 11 April, six days before PHAC reported 880 on 17 April. This example supports a local reporting advantage rather than a uniform annual lead.

\Needspace{8\baselineskip}
\noindent\textbf{Count agreement.} At reference report dates $\mathcal D$ with counts in both series, the mean absolute difference (MAD, cases) and weighted absolute percentage difference (WAPD) are
\begin{equation}
\mathrm{MAD}=\frac{\sum_{t\in\mathcal D}|W_t-R_t|}{|\mathcal D|},\qquad
\mathrm{WAPD}=100\,\frac{\sum_{t\in\mathcal D}|W_t-R_t|}{\sum_{t\in\mathcal D}R_t}.
\label{eq:period-count-agreement}
\end{equation}
These measures describe count agreement, not reporting order. Across 47 comparable reference dates, MAD was 56.49 cases and WAPD was 1.73\%; WAPD fell to 0.53\% in the second half, when all 26 dates had absolute relative differences within 5\%. The last workflow count was 5,359 on 29 December; PHAC reported 5,377~\cite{supp-phac_dec2025,supp-nysdoh_ghu} on 30 December with a 20 December cutoff.

\clearpage
\Needspace{16\baselineskip}\subsection{Provincial comparisons}

\begin{samepage}
Provincial comparisons evaluated the workflow series that contributed to the national totals. We selected Manitoba, Ontario and British Columbia for provincial comparisons. Provincial reference counts came from reported PHAC values or values uniquely determined from reported totals, increments and displayed proportions. Comparisons matched confirmed-only reference counts to workflow confirmed counts, and combined reference counts to combined or supported lower-bound workflow counts. Workflow values were missing at reference dates before the first workflow observation. Each annotated interval ends at the first report in the reference dataset reaching the threshold, so its length depends on which reference reports were available.

The Manitoba workflow series showed a sustained stepwise increase from spring through the end of the year (Fig.~\ref{fig:canada-manitoba}). Its series drew on 31 source observations, including 24 from Manitoba Health. Of 33 reference dates, 30 were comparable; the mean absolute difference was 2.43 cases and the weighted absolute percentage difference was 1.97\%; counts were equal at 17 dates. On the final reference date, 15 December, the workflow and reference counts were 301 and 300. The workflow reached 64 cases on 21 May; the first report in the reference dataset at or above this threshold was dated 7 July and reported 124 cases. The interval between these reports was 47 days.

\end{samepage}

\par\medskip\noindent\begin{minipage}{\linewidth}
\begin{center}\makebox[\linewidth][c]{\includegraphics[width=164mm]{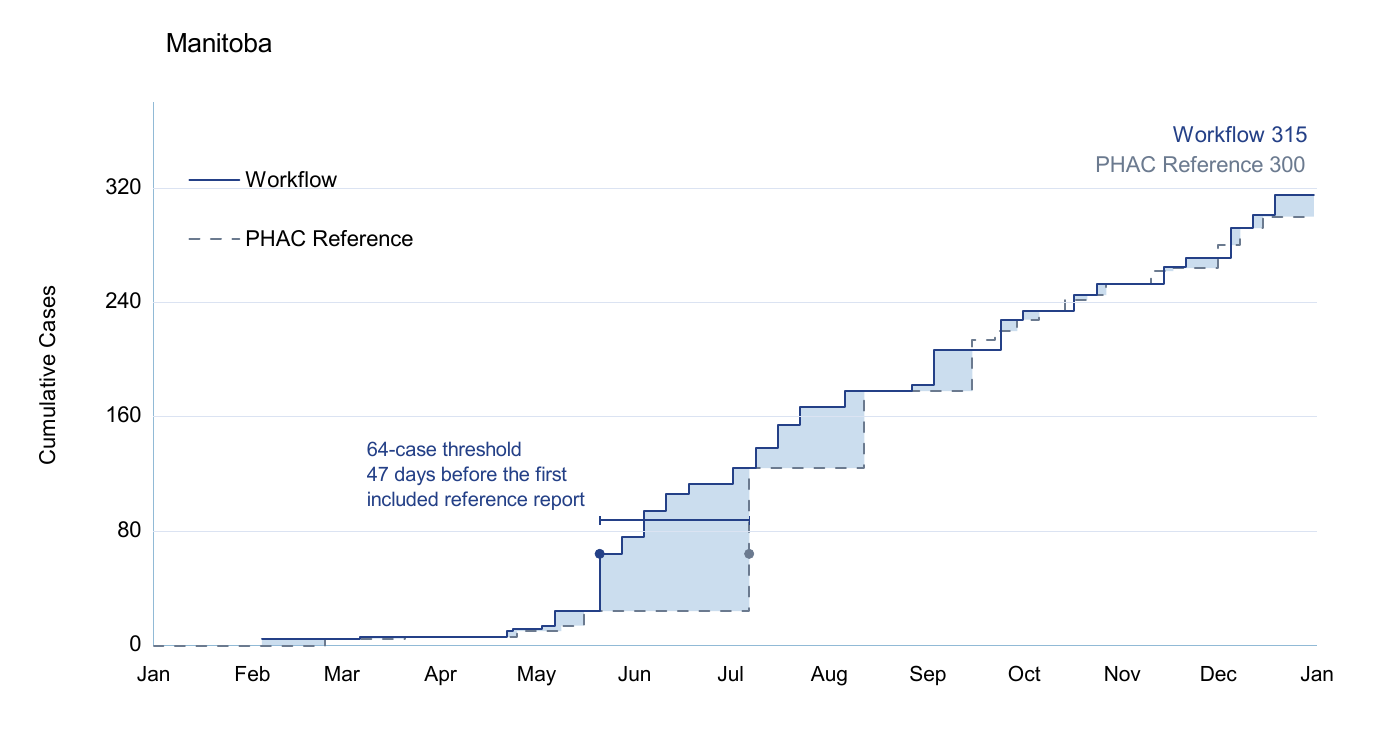}}\end{center}
\captionof{figure}{\textbf{Measles reporting in Manitoba.} Dark-blue solid and grey-blue dashed lines show Workflow and PHAC Reference. Pale-blue shading marks higher workflow counts. The workflow reached 64 cases on 21 May; the first included reference report at or above this threshold was dated 7 July (124 cases), 47 days later. Final counts are 315 (19 December) and 300 (15 December), respectively. Workflow observations begin on 4 February; earlier dates are left blank.}\label{fig:canada-manitoba}
\end{minipage}\par\medskip

\par\noindent\begin{minipage}{\linewidth}

\begin{samepage}
The Ontario workflow series increased most sharply in spring, then plateaued near 2,400 cases from late summer (Fig.~\ref{fig:canada-ontario}). Public Health Ontario supplied 39 of the 42 source observations. Of 42 reference dates, 38 were comparable, with a mean absolute difference of 34.29 cases and a weighted absolute percentage difference of 2.01\%. On 15 December, counts were 2,394 for the workflow and 2,395 for the reference. The workflow reached 1,453 cases on 8 May; the first report in the reference dataset at or above this threshold was dated 9 June and reported 2,046 cases. The interval between these reports was 32 days.

\end{samepage}

\par\medskip\noindent\begin{minipage}{\linewidth}
\begin{center}\makebox[\linewidth][c]{\includegraphics[width=164mm]{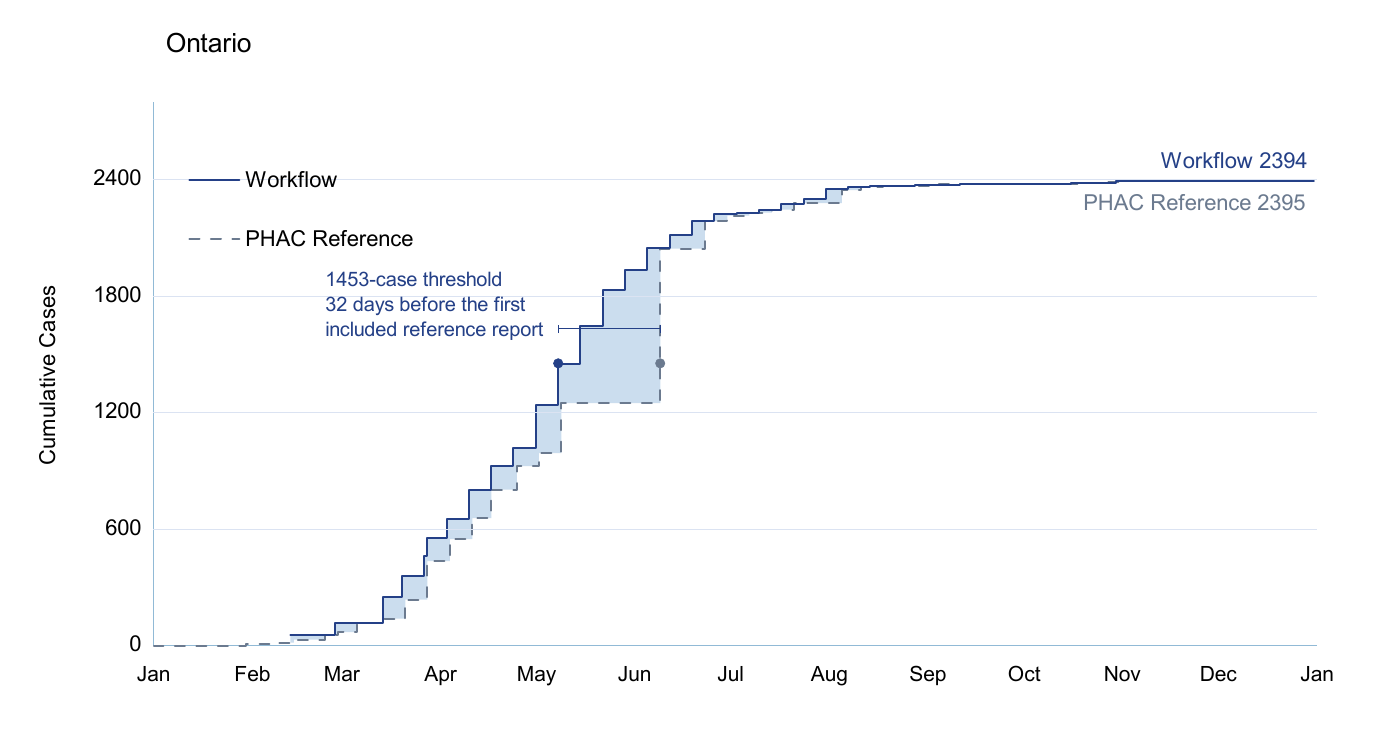}}\end{center}
\captionof{figure}{\textbf{Measles reporting in Ontario.} Lines and shading follow Fig.~\ref{fig:canada-manitoba}. The workflow reached 1,453 cases on 8 May; the first included reference report at or above this threshold was dated 9 June (2,046 cases), 32 days later. Final counts are 2,394 (23 December) and 2,395 (15 December), respectively. Workflow observations begin on 13 February; earlier dates are left blank.}\label{fig:canada-ontario}
\end{minipage}\par\medskip

\end{minipage}\par\medskip

\begin{samepage}
The British Columbia workflow series had relatively low cumulative counts in the first half of the year. A larger increase followed from summer through year end (Fig.~\ref{fig:canada-bc}). The evidence set contained 48 observations, including 39 from the BC Centre for Disease Control. Of 32 reference dates, 27 were comparable, with a mean absolute difference of 5.52 cases and a weighted absolute percentage difference of 3.27\%. On 15 December, the workflow and reference counts were 408 and 401. The workflow reached 173 cases on 5 August; the first report in the reference dataset at or above this threshold was dated 15 September and reported 268 cases. The interval between these reports was 41 days. The three provincial series thus extend the national comparison across different growth patterns and case count scales.

\end{samepage}

\par\medskip\noindent\begin{minipage}{\linewidth}
\begin{center}\makebox[\linewidth][c]{\includegraphics[width=164mm]{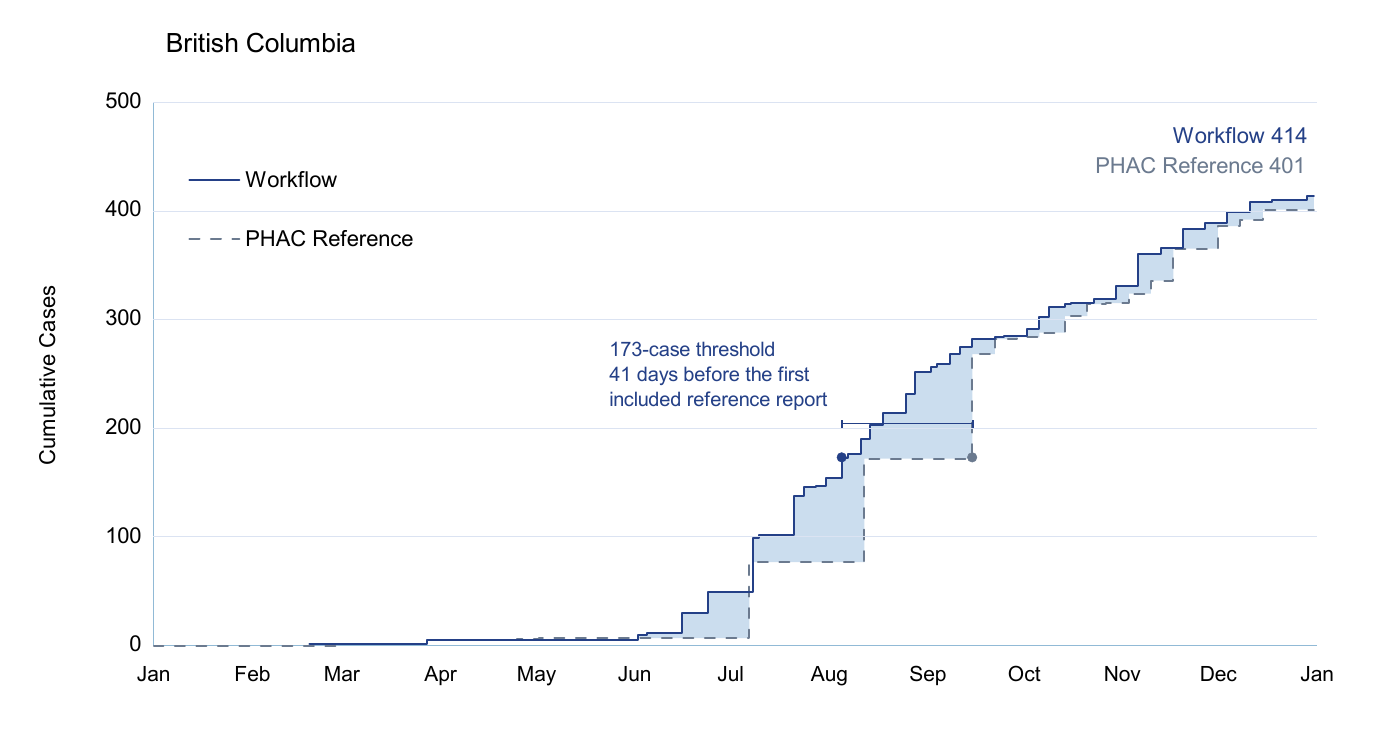}}\end{center}
\captionof{figure}{\textbf{Measles reporting in British Columbia.} Lines and shading follow Fig.~\ref{fig:canada-manitoba}. The workflow reached 173 cases on 5 August; the first included reference report at or above this threshold was dated 15 September (268 cases), 41 days later. Final counts are 414 (29 December) and 401 (15 December), respectively. Workflow observations begin on 19 February; earlier dates are left blank.}\label{fig:canada-bc}
\end{minipage}\par\medskip

\clearpage
\linespread{1}\selectfont
\section{Case study of mpox in Sierra Leone in 2025}\label{sec:mpox}

\Needspace{13\baselineskip}\subsection{Case overview and comparison design}

\begin{samepage}
We evaluated the workflow's source coverage, reporting continuity and count agreement for mpox in Sierra Leone during 2025. The analysis used cumulative confirmed counts at national scale, with a district-level comparison for Western Area Urban. National official publications, attributed news/media reports and regional institution reports supplied the workflow dataset. Publications from the World Health Organization (WHO)~\cite{supp-whompox,supp-whoafro} and institutional reports explicitly attributing their counts to WHO~\cite{supp-nysdoh_ghu} supplied the reference. The principal question was how media observations contributed to the coverage and agreement of the resulting reporting series.

\end{samepage}

\begin{samepage}
The national reference contained 31 eligible dated observations, which provided 24 reference comparison dates. Each series used publication dates when these were known. Otherwise, it used a date by which a report was known to be available. Statistical cutoffs were recorded separately. Between reports, the latest eligible value was carried forward; dates before the first observation remained missing. Repeated count statements were treated as duplicate information. Updates followed the ordering of statistical cutoffs, with decreases accepted when supported by a revision. Unresolved conflicts were retained outside the update sequence.

\end{samepage}

\Needspace{13\baselineskip}\subsection{Source composition and reporting channels}

\begin{samepage}
News/media constituted the largest source category in the national dataset (Fig.~\ref{fig:mpox-sources}a). Of 44 eligible observations, 27 came from news/media (61.4\%), eight from national official sources (18.2\%) and nine from a regional institution (20.5\%). The official sources comprised the National Public Health Agency (NPHA), the Sierra Leone News Agency (SLENA) and the Ministry of Information and Civic Education; all regional institution observations came from Africa CDC. This composition differed from the Canadian measles case study, in which official publications supplied 83.9\% of the retained provincial and territorial observations.

\end{samepage}

\begin{samepage}
The nine publishers provided complementary reporting channels (Fig.~\ref{fig:mpox-sources}b). Five were media outlets; Sierraloaded and ManoReporters contributed 18 and six observations, respectively, accounting for 24/27 media observations (88.9\%). Each media observation retained its stated attribution to government or national health authorities, linking the published count to the authority cited in the report.

\end{samepage}

\par\medskip\noindent\begin{minipage}{\linewidth}
\begin{center}\makebox[\linewidth][c]{\includegraphics[width=180mm]{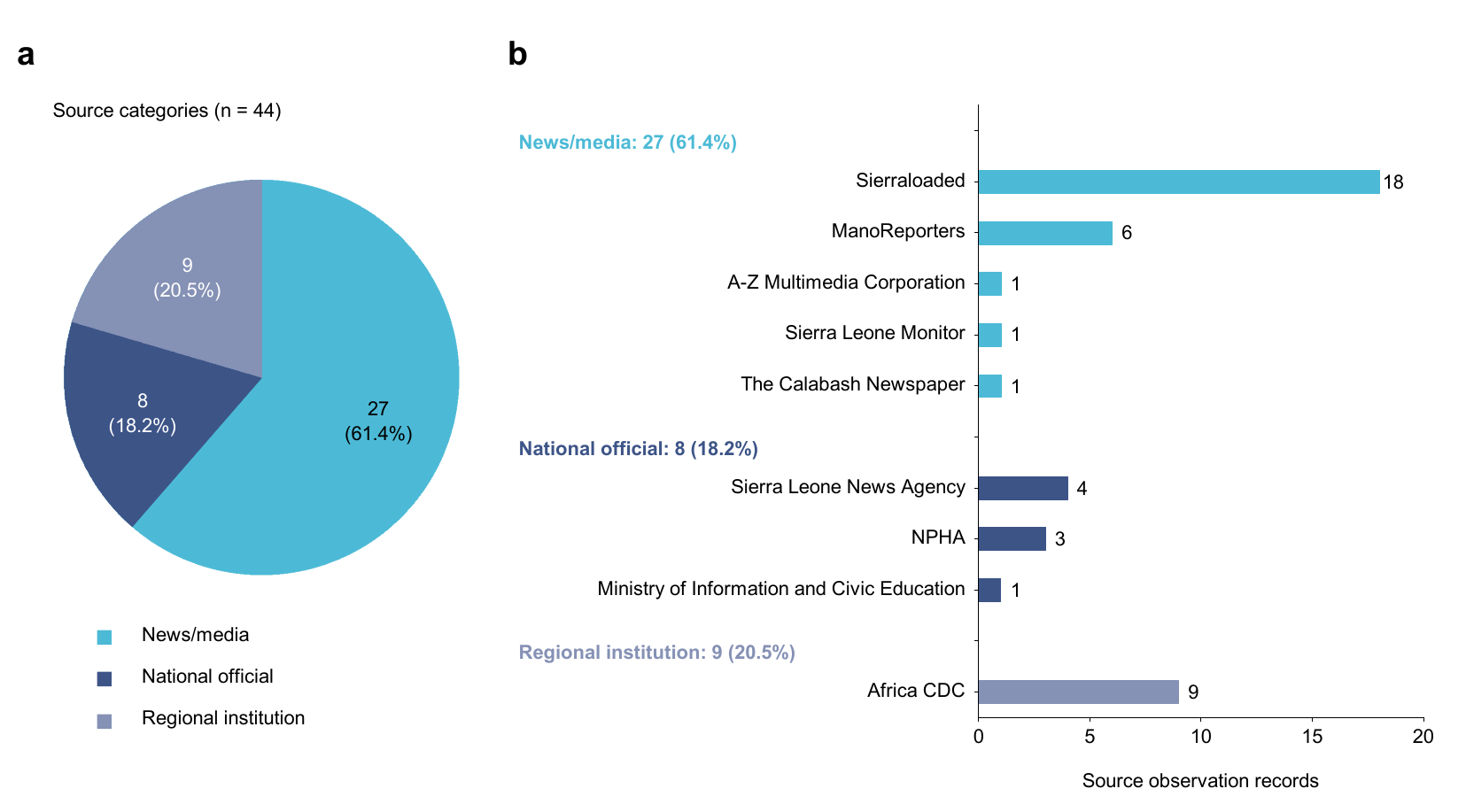}}\end{center}
\captionof{figure}{\textbf{Sources of national mpox observations.} a, Shares of the 44 eligible observations from news/media, national official sources and a regional institution. Slice labels show counts and percentages. b, Observation counts for all nine publishers, grouped by source category and ranked within each group. Each observation is counted once. Percentages sum to 100.1\% because of rounding. NPHA, National Public Health Agency; Africa CDC, Africa Centres for Disease Control and Prevention.}\label{fig:mpox-sources}
\end{minipage}\par\medskip

\begin{samepage}
The source categories contributed observations at different times of year (Table~\ref{tab:mpox-monthly-source}). Five of the six January observations were national official reports, whereas all 21 observations assigned to February--July came from media, including nine in May. Africa CDC supplied all seven observations in October--November. Combining these channels gave the workflow reporting coverage across different parts of the year.

\end{samepage}

\par\medskip\par\medskip

\par\medskip\begingroup\small\renewcommand{\arraystretch}{1.12}
\begin{longtable}{@{}lrrrr@{}}
\caption{Sierra Leone mpox: monthly source observations.}\label{tab:mpox-monthly-source}\\
\toprule
Month & National official & Media & Regional & Total \\
\midrule\endfirsthead
\toprule
Month & National official & Media & Regional & Total \\
\midrule\endhead
Jan & 5 & 1 & 0 & 6 \\
Feb & 0 & 1 & 0 & 1 \\
Mar & 0 & 2 & 0 & 2 \\
Apr & 0 & 4 & 0 & 4 \\
May & 0 & 9 & 0 & 9 \\
Jun & 0 & 2 & 0 & 2 \\
Jul & 0 & 3 & 0 & 3 \\
Aug & 1 & 4 & 0 & 5 \\
Sep & 1 & 1 & 2 & 4 \\
Oct & 0 & 0 & 4 & 4 \\
Nov & 0 & 0 & 3 & 3 \\
Dec & 1 & 0 & 0 & 1 \\
\midrule
Total & 8 & 27 & 9 & 44 \\
\bottomrule\end{longtable}\endgroup

Each retained observation is counted once. Months use recorded public availability dates, including documented upper bounds when necessary.

\Needspace{10\baselineskip}\subsection{Media contributions to the reconstructed reporting series}

\begin{samepage}
We assessed media contributions by comparing the complete 44-observation dataset with the 17-observation subset comprising eight national official and nine Africa CDC observations. Both source sets used the same sequence-construction rules, so the comparison measured changes in the reconstructed reporting series when media observations were included.

\end{samepage}

\begin{samepage}
Media observations improved reporting continuity. Report age was the number of days since the latest retained report's publication date or, when this was uncertain, its documented available-by date. A later accepted report reset report age even when the cumulative count was unchanged; this measure describes reporting recency rather than the time since the last count change or the statistical cutoff. Including media increased the number of distinct dates with retained reports from 17 to 43. We assessed report age over the same 355 calendar days, from 11 January to 31 December. Without media, a report no more than seven days old was available on 100 days (28.2\%). With all sources, this increased to 233 days (65.6\%; Fig.~\ref{fig:mpox-media}a). Days on which the latest report was more than 28 days old fell from 169 (47.6\%) to five (1.4\%), and median report age fell from 24 to five days.

\end{samepage}

\begin{samepage}
Including media also brought the workflow counts closer to the reference at most comparison dates (Fig.~\ref{fig:mpox-media}b). At the same 24 reference comparison dates, the absolute count difference decreased at 18 dates (75.0\%), was unchanged at four (16.7\%) and increased at two (8.3\%). The mean absolute count difference fell from 1,374.63 to 162.63 cases.

\end{samepage}

\par\vspace{-6pt}\noindent\begin{minipage}{\linewidth}
\enlargethispage{10pt}
\begin{center}\makebox[\linewidth][c]{\includegraphics[width=180mm]{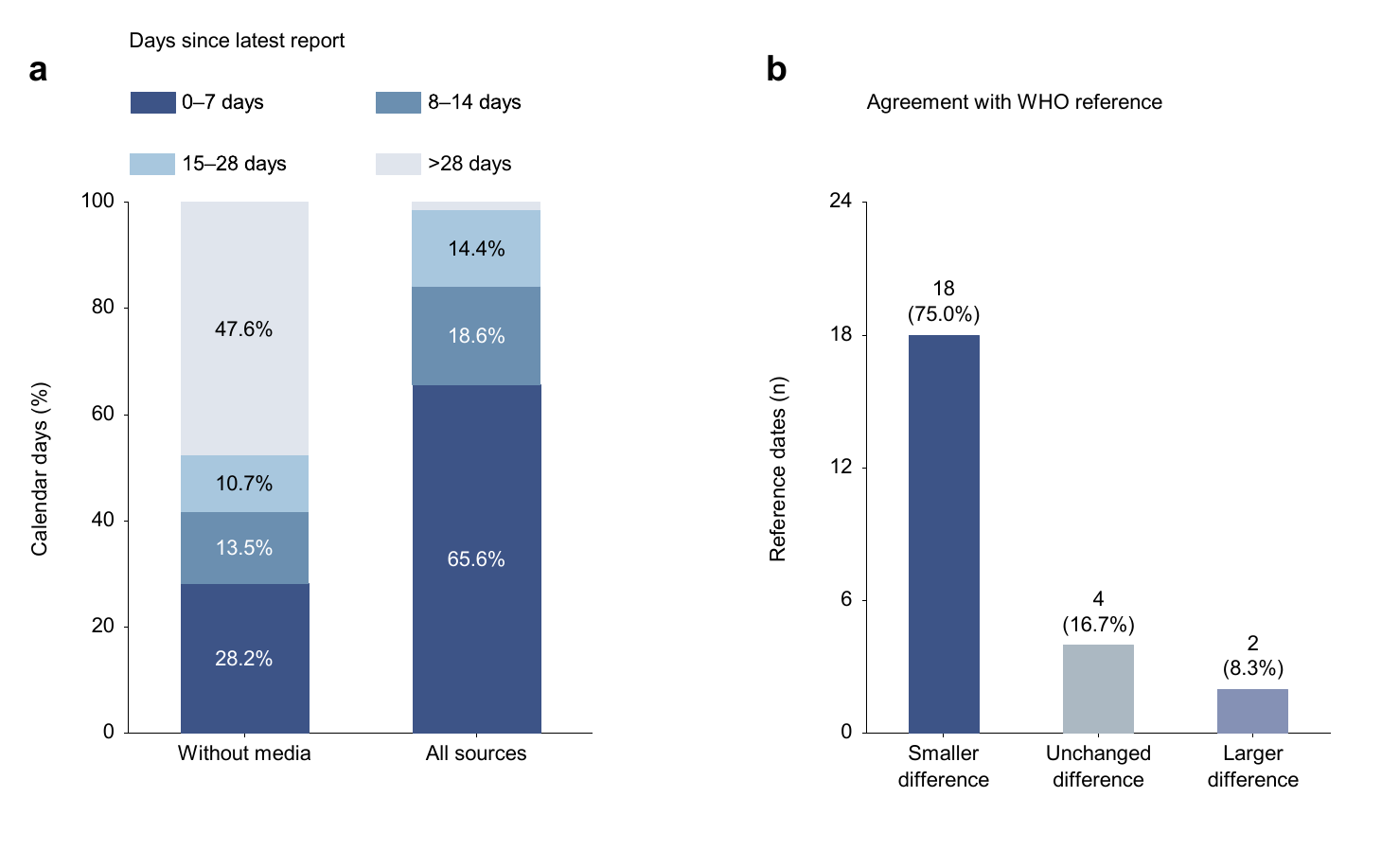}}\end{center}\vspace{-32pt}
\captionof{figure}{\textbf{Effects of media reports on reporting continuity and agreement with the reference.} Both groups include eight national official and nine regional institution observations; the all-source group also includes 27 media observations. a, Percentage of calendar days grouped by days since the latest included report, across the same 355 days (11 January--31 December 2025). The unlabelled top segment of the all-source bar represents five days (1.4\%) with no report in the preceding 28 days. b, Changes in the absolute count difference, $|W-R|$, after adding media reports, compared at the same 24 reference dates. Percentages are based on these 24 dates. Equal absolute differences can occur with different workflow counts. These comparisons use report dates and measure agreement between public-reporting series.}\label{fig:mpox-media}
\end{minipage}\par\medskip

\clearpage
\subsection{Monthly reporting continuity and media contribution}

For mpox, we evaluated report freshness over the shared 355-day window from 11 January to 31 December 2025. January therefore contributes 21 days; the ten days before the first report were excluded from both series. Let $L_s(t)$ be the latest retained reporting date at or before day $t$ for source set $s$, and let $D_p$ be the shared calendar days in period $p$. The fraction of days with a report no more than seven days old is

\begin{equation}
F_{s,p}^{(7)}=\frac{100}{|D_p|}\sum_{t\in D_p}\mathbf{1}\{t-L_s(t)\leq7\ \mathrm{days}\}.
\label{eq:period-freshness}
\end{equation}

All sources and the subset without media were evaluated on exactly the same days with the same carry-forward rule. Media increased this fraction in February--August and October, and left it unchanged in January, September, November and December (Table~\ref{tab:monthly-freshness}; Fig.~\ref{fig:monthly-freshness}). In May, the complete series supplied a report no more than seven days old on all 31 days, versus none without media. The first-half gain was 89 days, or 52.0 percentage points (64.3\% versus 12.3\%); the second-half gain was 44 days, or 23.9 points (66.8\% versus 42.9\%). These contrasts locate the largest continuity contribution in the first half, while retaining the later improvements.

The October gain reflects the recency of reports carried forward from September, rather than new October media observations: all four retained October observations came from Africa CDC. Across the full year, the complete series yielded 233/355 fresh-report days (65.6\%) versus 100/355 (28.2\%), and median report age decreased from 24 to five days.

\begingroup\small\setlength{\tabcolsep}{4pt}\renewcommand{\arraystretch}{1.15}
\begin{longtable}{@{}>{\raggedright\arraybackslash}p{23mm}rrrr@{}}
\caption{Mpox report freshness on the shared calendar, by month and half-year.}\label{tab:monthly-freshness}\\
\toprule
Period & \shortstack{Shared\\days} & \shortstack{All sources\\days (\%)} & \shortstack{Without media\\days (\%)} & \shortstack{Gain\\(pp)} \\
\midrule\endfirsthead
\toprule
Period & \shortstack{Shared\\days} & \shortstack{All sources\\days (\%)} & \shortstack{Without media\\days (\%)} & \shortstack{Gain\\(pp)} \\
\midrule\endhead
Jan & 21 & 21 (100.0) & 21 (100.0) & 0.0 \\
Feb & 28 & 10 (35.7) & 0 (0.0) & 35.7 \\
Mar & 31 & 16 (51.6) & 0 (0.0) & 51.6 \\
Apr & 30 & 14 (46.7) & 0 (0.0) & 46.7 \\
May & 31 & 31 (100.0) & 0 (0.0) & 100.0 \\
Jun & 30 & 18 (60.0) & 0 (0.0) & 60.0 \\
Jul & 31 & 21 (67.7) & 0 (0.0) & 67.7 \\
Aug & 31 & 24 (77.4) & 8 (25.8) & 51.6 \\
Sep & 30 & 18 (60.0) & 18 (60.0) & 0.0 \\
Oct & 31 & 31 (100.0) & 24 (77.4) & 22.6 \\
Nov & 30 & 21 (70.0) & 21 (70.0) & 0.0 \\
Dec & 31 & 8 (25.8) & 8 (25.8) & 0.0 \\
\midrule
H1 & 171 & 110 (64.3) & 21 (12.3) & 52.0 \\
H2 & 184 & 123 (66.8) & 79 (42.9) & 23.9 \\
\midrule
Annual & 355 & 233 (65.6) & 100 (28.2) & 37.5 \\
\bottomrule
\end{longtable}
\endgroup

Days (\%) counts calendar days on which report age was at most seven days; pp denotes percentage points. Shared-day denominators are identical for both source sets within each row. Daily records and source-composition tables retain the report dates supporting each value.

\clearpage
\noindent\begin{minipage}{\linewidth}
\begin{center}\makebox[\linewidth][c]{\includegraphics[width=164mm]{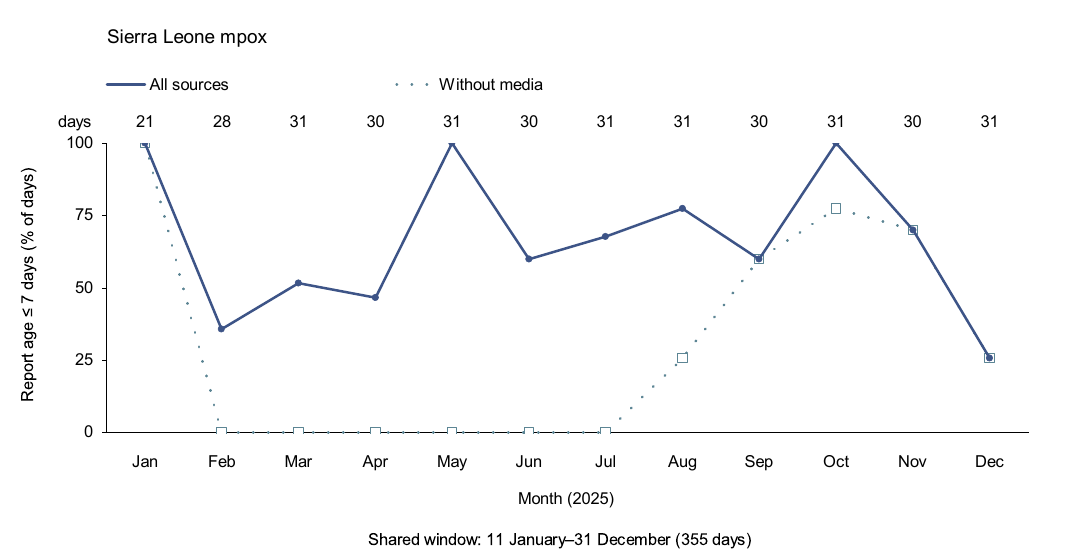}}\end{center}
\captionof{figure}{\textbf{Monthly contribution of media to mpox report freshness.} All-source and without-media series are compared across every month of 2025 on the same daily calendar. Values are the percentage of days on which the most recent retained report was no more than seven days old. January includes 11--31 January; all other months use every calendar day. The common annual window comprises 355 days. Recency is carried forward from the most recent report, so a monthly difference may reflect a report published in the preceding month. Table~\ref{tab:monthly-freshness} gives exact numerators, denominators and half-year summaries. These descriptive daily summaries are temporally dependent.}\label{fig:monthly-freshness}
\end{minipage}

\clearpage

\Needspace{8\baselineskip}
\subsection{Duration and magnitude of reported-count advantage}\label{sec:mpox-count-advantage}

Daily duration and the proportion of days with higher workflow counts use Eq.~\ref{eq:reported-count-lead}; mean and relative count differences use Eqs.~\ref{eq:reported-count-lead} and \ref{eq:relative-count-difference}. We apply the same documented availability dates, carry-forward rule and missing-data treatment as in the national comparison. These measures describe the duration and magnitude of higher reported counts, separately from the threshold crossing lead.

Both national series had comparable counts from 1 February onward: 334 days, rather than the 355-day workflow freshness window. Workflow counts were higher on 178/334 days (53.3\%), equal on 56 and lower on 100; the mean signed daily difference was +160.47 cases, corresponding to a relative difference of +4.87\%. May had 31/31 higher-count days with a mean of +1,073.90 cases, and October had 31/31 with a mean of +92.84. In the first six tracking months, 97/160 comparable days (60.6\%) had higher counts, with a mean difference of +283.19 cases. The first nine months had 152/252 higher-count days (60.3\%), a mean difference of +199.75 cases and a relative difference of +7.66\%. These periods contrast with March (7/31 higher-count days; mean -2.74) and December (all 31 days equal).

The first tracking month contains only 10 comparable days out of 31. Its 100\% higher-count fraction therefore means 10/10 comparable days, not 31/31 days. The calendar year contains 31 unavailable days; higher-count days occupy 48.8\% of all 365 days and 53.3\% of the 334 comparable days. Carry-forward states describe retained information, including periods without a new reference release. The separate freshness analysis quantifies how recently each source updated.

Figure~\ref{fig:mpox-window-metrics} summarises these monthly and tracking-window measures; Tables~\ref{tab:mpox-leading-case-monthly} and \ref{tab:mpox-leading-case-windows} provide exact values and coverage. All panels include the same available-by observations. The descriptive days and overlapping windows are not independent replicates.

\clearpage\newgeometry{left=15mm,right=15mm,top=18mm,bottom=18mm,headheight=14pt}
\begin{center}\includegraphics[width=180mm]{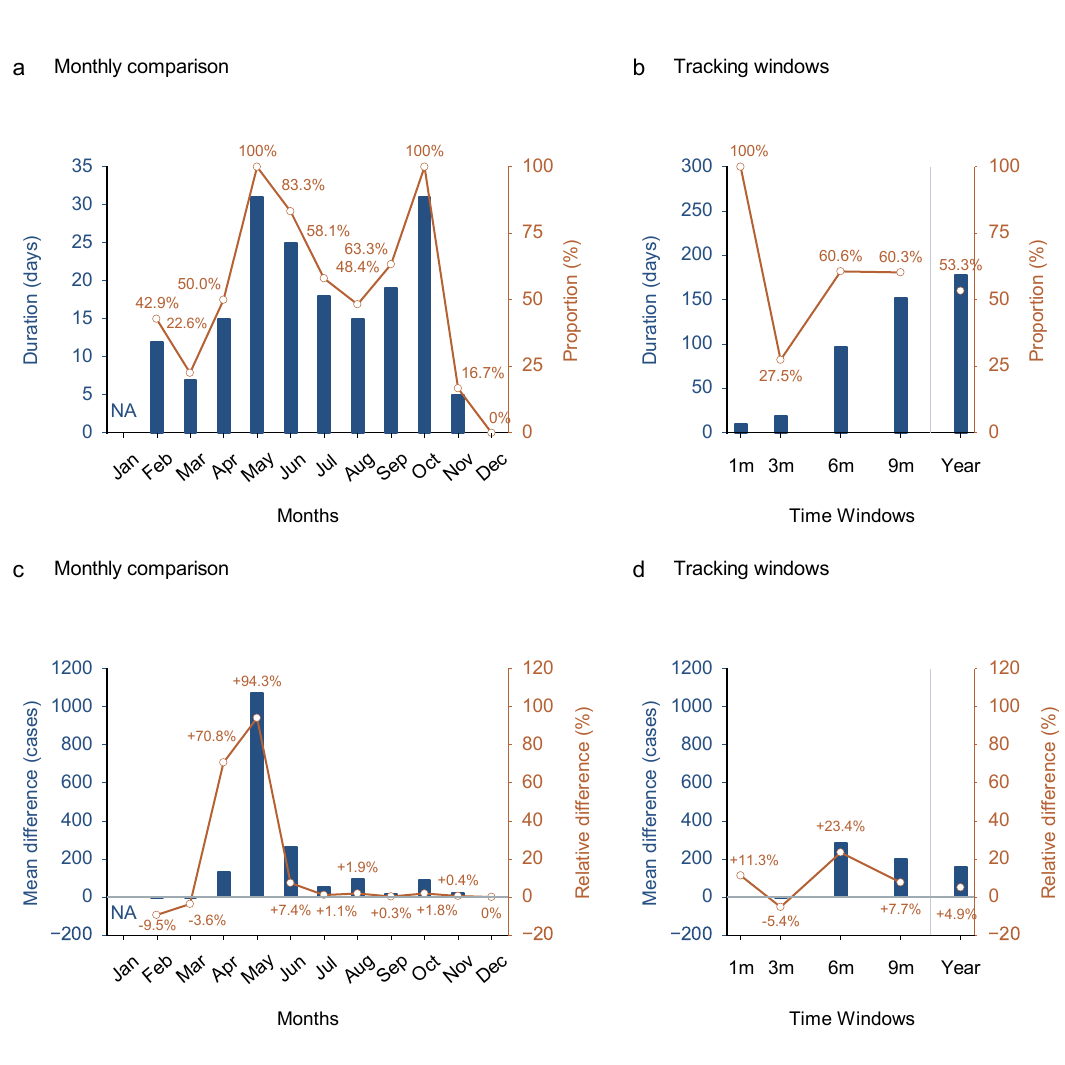}\end{center}
\captionof{figure}{\textbf{Duration and magnitude of reported-count advantage for Sierra Leone mpox.} a,b, Days with $W>R$ (blue bars, left axis) and their proportion among comparable days (orange line, right axis). c,d, Mean signed $W-R$ (blue bars, left axis) and relative difference $\rho$ (orange line, right axis; Eq.~\ref{eq:relative-count-difference}). a,c, Calendar months. b,d, First one, three, six and nine tracking months, with Year shown separately. Only orange percentages are labelled. Positive differences indicate higher workflow-reported counts. NA denotes no comparable days; zero is an observed value. Duration is not mean reporting-date advance.}\label{fig:mpox-window-metrics}
\loadgeometry{si}\clearpage

\begingroup\small\setlength{\tabcolsep}{4pt}\renewcommand{\arraystretch}{1.15}
\begin{longtable}{@{}>{\raggedright\arraybackslash}p{27mm}>{\centering\arraybackslash}p{26mm}>{\centering\arraybackslash}p{30mm}>{\centering\arraybackslash}p{18mm}>{\centering\arraybackslash}p{22mm}>{\centering\arraybackslash}p{23mm}@{}}
\caption{Sierra Leone mpox: daily reported-count advantage by calendar period.}\label{tab:mpox-leading-case-monthly}\\
\toprule
Period & \shortstack{Comparable /\\calendar days} & \shortstack{Higher / equal\\/ lower days} & \shortstack{Higher\\(\%)} & \shortstack{Mean $D$\\(cases)} & \shortstack{Relative $\rho$\\(\%)} \\
\midrule\endfirsthead
\toprule
Period & \shortstack{Comparable /\\calendar days} & \shortstack{Higher / equal\\/ lower days} & \shortstack{Higher\\(\%)} & \shortstack{Mean $D$\\(cases)} & \shortstack{Relative $\rho$\\(\%)} \\
\midrule\endhead
Jan & 0/31 & --- & --- & --- & --- \\
Feb & 28/28 & 12/0/16 & 42.9 & -1.89 & -9.55 \\
Mar & 31/31 & 7/0/24 & 22.6 & -2.74 & -3.62 \\
Apr & 30/30 & 15/0/15 & 50.0 & +131.43 & +70.78 \\
May & 31/31 & 31/0/0 & 100.0 & +1073.90 & +94.26 \\
Jun & 30/30 & 25/0/5 & 83.3 & +262.67 & +7.41 \\
Jul & 31/31 & 18/0/13 & 58.1 & +52.71 & +1.15 \\
Aug & 31/31 & 15/0/16 & 48.4 & +95.39 & +1.91 \\
Sep & 30/30 & 19/0/11 & 63.3 & +14.70 & +0.28 \\
Oct & 31/31 & 31/0/0 & 100.0 & +92.84 & +1.75 \\
Nov & 30/30 & 5/25/0 & 16.7 & +23.67 & +0.44 \\
Dec & 31/31 & 0/31/0 & 0.0 & 0.00 & 0.00 \\
Year & 334/365 & 178/56/100 & 53.3 & +160.47 & +4.87 \\
\bottomrule
\end{longtable}
\endgroup

The higher-day percentage divides by comparable days; relative $\rho$ divides the mean signed difference by the reference mean over the same dates (Eq.~\ref{eq:relative-count-difference}). Both count difference measures retain positive, zero and negative daily differences. A dash denotes no eligible comparison or a zero reference denominator; observed zero is reported as zero. Year retains 365 days in the coverage column.

\clearpage
\begingroup\small\setlength{\tabcolsep}{4pt}\renewcommand{\arraystretch}{1.15}
\begin{longtable}{@{}>{\raggedright\arraybackslash}p{27mm}>{\centering\arraybackslash}p{26mm}>{\centering\arraybackslash}p{30mm}>{\centering\arraybackslash}p{18mm}>{\centering\arraybackslash}p{22mm}>{\centering\arraybackslash}p{23mm}@{}}
\caption{Sierra Leone mpox: daily reported-count advantage by tracking window.}\label{tab:mpox-leading-case-windows}\\
\toprule
Period & \shortstack{Comparable /\\calendar days} & \shortstack{Higher / equal\\/ lower days} & \shortstack{Higher\\(\%)} & \shortstack{Mean $D$\\(cases)} & \shortstack{Relative $\rho$\\(\%)} \\
\midrule\endfirsthead
\toprule
Period & \shortstack{Comparable /\\calendar days} & \shortstack{Higher / equal\\/ lower days} & \shortstack{Higher\\(\%)} & \shortstack{Mean $D$\\(cases)} & \shortstack{Relative $\rho$\\(\%)} \\
\midrule\endhead
First month & 10/31 & 10/0/0 & 100.0 & +1.80 & +11.25 \\
First 3 months & 69/90 & 19/0/50 & 27.5 & -3.16 & -5.38 \\
First 6 months & 160/181 & 97/0/63 & 60.6 & +283.19 & +23.37 \\
First 9 months & 252/273 & 152/0/100 & 60.3 & +199.75 & +7.66 \\
Year & 334/365 & 178/56/100 & 53.3 & +160.47 & +4.87 \\
\bottomrule
\end{longtable}
\endgroup

The higher-day percentage divides by comparable days; relative $\rho$ divides the mean signed difference by the reference mean over the same dates (Eq.~\ref{eq:relative-count-difference}). Both count difference measures retain positive, zero and negative daily differences. A dash denotes no eligible comparison or a zero reference denominator; observed zero is reported as zero. Year retains 365 days in the coverage column.

Tracking windows start on 11 January 2025. The first one, three, six and nine months end on 10 February, 10 April, 10 July and 10 October, respectively. The nested windows overlap; Year covers the full calendar year, 1 January--31 December.

Using the first-crossing and date-interval method in Section~\ref{sec:canada-count-advantage}, the workflow was earlier at 9/35 national thresholds (25.7\%), later at 16 and undetermined at ten. The 15 exact date pairs had a mean reference-minus-workflow difference of $-0.93$ days. At 2,648 cases, the Sierraloaded report dated 20 May preceded the WHO report dated 29 May by nine days; their stated statistical cutoffs were 19 and 18 May, respectively. These threshold results describe reporting-date differences, whereas the daily measures above quantify how long and by how much reported counts were higher.

At the 24 reference comparison dates, the weighted absolute percentage difference was 6.56\%, and absolute relative differences were within 5\% at 10/24 dates (41.7\%), including all nine dates from July onward. Both national series ended at 5,442 cases: the workflow retained this count in Africa CDC's 6 November report and SLENA's 15 December report, while the 18 December reference report cited WHO counts with a 7 December statistical cutoff.

\clearpage
\renewcommand{\bibsection}{\section*{Supplementary References}}
\phantomsection

{\small
\begin{thebibliography}{99}

\bibitem{cdc}
German, R. R. et al. Updated guidelines for evaluating public health surveillance systems: recommendations from the Guidelines Working Group. \href{https://www.cdc.gov/mmwr/preview/mmwrhtml/rr5013a1.htm}{\textit{MMWR Recomm. Rep.} \textbf{50}, 1--35 (2001)}.

\bibitem{langmuir1971}
Langmuir, A. D. Communicable disease surveillance: evolution of the concept of surveillance in the United States. \href{https://doi.org/10.1177/003591577106400646}{\textit{Proc. R. Soc. Med.} \textbf{64}, 681--684 (1971)}.

\bibitem{lee2011}
Lee, L. M. \& Thacker, S. B. The cornerstone of public health practice: public health surveillance, 1961--2011. \href{https://www.cdc.gov/mmwr/preview/mmwrhtml/su6004a4.htm}{\textit{MMWR Suppl.} \textbf{60}, 15--21 (2011)}.

\bibitem{snow}
Snow, J. \href{https://epi-snow.ph.ucla.edu/Stream2_BSPoutbreak_a.html}{\textit{On the Mode of Communication of Cholera}} 2nd edn, 38--55 (John Churchill, 1855).

\bibitem{gphin}
Mawudeku, A. \& Blench, M. Global Public Health Intelligence Network (GPHIN). In \href{https://aclanthology.org/2005.mtsummit-invited.4/}{\textit{Proceedings of Machine Translation Summit X: Invited Papers}} (Phuket, Thailand, 2005).

\bibitem{epispider}
Tolentino, H. et al. Scanning the emerging infectious diseases horizon---visualizing ProMED emails using EpiSPIDER. \href{https://lhncbc.nlm.nih.gov/LHC-publications/pubs/ScanningtheEmergingInfectiousDiseasesHorizonVisualizingProMEDEmailsUsingEpiSPIDER.html}{\textit{Adv. Dis. Surveill.} \textbf{2}, 169 (2007)}.

\bibitem{healthmap}
Freifeld, C. C., Mandl, K. D., Reis, B. Y. \& Brownstein, J. S. HealthMap: global infectious disease monitoring through automated classification and visualization of Internet media reports. \href{https://doi.org/10.1197/jamia.M2544}{\textit{J. Am. Med. Inform. Assoc.} \textbf{15}, 150--157 (2008)}.

\bibitem{dong2020}
Dong, E., Du, H. \& Gardner, L. An interactive web-based dashboard to track COVID-19 in real time. \href{https://doi.org/10.1016/S1473-3099(20)30120-1}{\textit{Lancet Infect. Dis.} \textbf{20}, 533--534 (2020)}.

\bibitem{dong2022}
Dong, E. et al. The Johns Hopkins University Center for Systems Science and Engineering COVID-19 Dashboard: data collection process, challenges faced, and lessons learned. \href{https://doi.org/10.1016/S1473-3099(22)00434-0}{\textit{Lancet Infect. Dis.} \textbf{22}, e370--e376 (2022)}.

\bibitem{timeliness}
Espinosa, L., Altunina, O. \& Salathé, M. Timeliness of online COVID-19 reports from official sources. \href{https://doi.org/10.3389/fpubh.2022.1027812}{\textit{Front. Public Health} \textbf{10}, 1027812 (2023)}.

\bibitem{harrod}
Harrod, K. K., Bhandari, P. \& Anastasopoulos, A. From text to maps: LLM-driven extraction and geotagging of epidemiological data. In \href{https://doi.org/10.18653/v1/2024.nlp4pi-1.24}{\textit{Proceedings of the Third Workshop on NLP for Positive Impact} 258--270} (Association for Computational Linguistics, 2024).

\bibitem{pandem}
Tighe, C. et al. Building and developing a tool (PANDEM-2 dashboard) to strengthen pandemic management: participatory design study. \href{https://doi.org/10.2196/52119}{\textit{JMIR Public Health Surveill.} \textbf{11}, e52119 (2025)}.

\bibitem{kraemer2025}
Kraemer, M. U. G. et al. Artificial intelligence for modelling infectious disease epidemics. \href{https://doi.org/10.1038/s41586-024-08564-w}{\textit{Nature} \textbf{638}, 623--635 (2025)}.

\bibitem{sentinel}
Pant, D. et al. Health Sentinel: an AI pipeline for real-time disease outbreak detection. In \href{https://doi.org/10.18653/v1/2025.nlp4pi-1.3}{\textit{Proceedings of the Fourth Workshop on NLP for Positive Impact (NLP4PI)} 23--42} (Association for Computational Linguistics, 2025).

\bibitem{xu}
Xu, B. et al. Epidemiological data from the COVID-19 outbreak, real-time case information. \href{https://doi.org/10.1038/s41597-020-0448-0}{\textit{Sci. Data} \textbf{7}, 106 (2020)}.

\bibitem{mpox}
Kraemer, M. U. G. et al. Tracking the 2022 monkeypox outbreak with epidemiological data in real-time. \href{https://doi.org/10.1016/S1473-3099(22)00359-0}{\textit{Lancet Infect. Dis.} \textbf{22}, 941--942 (2022)}.

\bibitem{hed}
Badker, R. et al. Constructing a global human epidemic database using open-source digital biosurveillance. \href{https://doi.org/10.1038/s41597-025-04663-z}{\textit{Sci. Data} \textbf{12}, 344 (2025)}.

\bibitem{eioskorea}
Kim, S. et al. Event-based surveillance in the Republic of Korea: assessment of the effectiveness of Epidemic Intelligence from Open Sources. \href{https://doi.org/10.5365/wpsar.2025.16.3.1151}{\textit{Western Pac. Surveill. Response J.} \textbf{16}, 1--7 (2025)}.

\bibitem{epicore}
Divi, N. et al. Using EpiCore to enable rapid verification of potential health threats: illustrated use cases and summary statistics. \href{https://doi.org/10.2196/52093}{\textit{JMIR Public Health Surveill.} \textbf{10}, e52093 (2024)}.

\bibitem{eiosafrica}
Williams, G. S. et al. Evaluation of the epidemic intelligence from open sources (EIOS) system for the early detection of outbreaks and health emergencies in the African region. \href{https://doi.org/10.1186/s12889-025-21998-9}{\textit{BMC Public Health} \textbf{25}, 857 (2025)}.

\bibitem{toolformer}
Schick, T. et al. Toolformer: language models can teach themselves to use tools. \href{https://doi.org/10.52202/075280-2997}{\textit{Adv. Neural Inf. Process. Syst.} \textbf{36}, 68539--68551 (2023)}.

\bibitem{react}
Yao, S. et al. ReAct: synergizing reasoning and acting in language models. In \href{https://openreview.net/forum?id=WE_vluYUL-X}{\textit{International Conference on Learning Representations}} (2023).

\bibitem{reflexion}
Shinn, N., Cassano, F., Gopinath, A., Narasimhan, K. \& Yao, S. Reflexion: language agents with verbal reinforcement learning. \href{https://doi.org/10.52202/075280-0377}{\textit{Adv. Neural Inf. Process. Syst.} \textbf{36}, 8634--8652 (2023)}.

\bibitem{autogen}
Wu, Q. et al. AutoGen: enabling next-gen LLM applications via multi-agent conversations. In \href{https://openreview.net/forum?id=BAakY1hNKS}{\textit{First Conference on Language Modeling}} (2024).

\bibitem{consoli2025}
Consoli, S. et al. An epidemiological knowledge graph extracted from the World Health Organization's Disease Outbreak News. \href{https://doi.org/10.1038/s41597-025-05276-2}{\textit{Sci. Data} \textbf{12}, 970 (2025)}.

\bibitem{yang2026}
Yang, L. et al. AI agents and epidemic intelligence on respiratory infectious diseases: toward a conceptual framework integrating decision support. \href{https://doi.org/10.2196/86936}{\textit{J. Med. Internet Res.} \textbf{28}, e86936 (2026)}.

\bibitem{bhadelia2026}
Bhadelia, N., Paschalidis, I. C., Brownstein, J. S. \& Lassman, B. A BEACON for novel disease threats: leveraging artificial intelligence for informal event-based outbreak surveillance. \href{https://doi.org/10.1093/infdis/jiaf642}{\textit{J. Infect. Dis.} \textbf{233}, e287--e289 (2026)}.

\bibitem{globalhealth2026}
Global.health. \textit{Hondius hantavirus 2026 outbreak data}. \href{https://github.com/kraemer-lab/Hondius_hantavirus_h2026/tree/16dd94e127223a3abc5295cf72c649c132b53509}{GitHub}, version of 30 June 2026.

\bibitem{whohantavirus2026}
World Health Organization. \href{https://www.who.int/emergencies/disease-outbreak-news/item/2026-DON604}{\textit{Hantavirus outbreak linked to cruise ship travel, Multi-locations}}. Disease Outbreak News (28 May 2026).

\bibitem{phac_dec2025}
Public Health Agency of Canada. \href{https://health-infobase.canada.ca/measles-rubella/archive/2025/}{\textit{Measles and Rubella Weekly Monitoring Report}}. Health Infobase (2025).

\bibitem{tavily}
Tavily. \href{https://docs.tavily.com/documentation/api-reference/endpoint/search}{\textit{Tavily Search API}}. API documentation (2026).

\bibitem{si-eurosurveillance}
van den Berg, O. E. et al. Andes virus outbreak linked to expedition cruise ship travel, multi-country investigation and response, April to June 2026. \href{https://doi.org/10.2807/1560-7917.ES.2026.31.24.2600477}{\textit{Euro Surveill.} \textbf{31}, 2600477 (2026)}.

\bibitem{si-whorra}
World Health Organization. \href{https://www.who.int/publications/m/item/who-rapid-risk-assessment---hantavirus-outbreak-caused-by-andes-virus--global-v.2}{\textit{Hantavirus outbreak caused by Andes virus, Global v.2}}. WHO Rapid Risk Assessment (17 May 2026).

\end{thebibliography}
}


\addcontentsline{toc}{section}{Supplementary References}
{\small
\begin{thebibliography}{99}

\bibitem{supp-langflow}
Langflow. \href{https://github.com/langflow-ai/langflow}{\textit{Langflow}}. GitHub (2026).

\bibitem{supp-langgraph}
LangChain. \href{https://github.com/langchain-ai/langgraph}{\textit{LangGraph}}. GitHub (2026).

\bibitem{supp-claudesonnet46}
Anthropic. \href{https://www.anthropic.com/claude-sonnet-4-6-system-card}{\textit{System Card: Claude Sonnet 4.6}}. Model documentation (17 February 2026).

\bibitem{supp-tavily}
Tavily. \href{https://docs.tavily.com/documentation/api-reference/endpoint/search}{\textit{Tavily Search API}}. API documentation (2026).

\bibitem{supp-tavilyextract}
Tavily. \href{https://docs.tavily.com/documentation/api-reference/endpoint/extract}{\textit{Tavily Extract API}}. API documentation (2026).

\bibitem{supp-pydantic}
Colvin, S. et al. \href{https://github.com/pydantic/pydantic}{\textit{Pydantic Validation}}. GitHub (2026).

\bibitem{supp-globalhealth2026}
Global.health. \textit{Hondius hantavirus 2026 outbreak data}. \href{https://github.com/kraemer-lab/Hondius_hantavirus_h2026/tree/16dd94e127223a3abc5295cf72c649c132b53509}{GitHub}, version of 30 June 2026.

\bibitem{supp-whodon599}
World Health Organization. \href{https://www.who.int/emergencies/disease-outbreak-news/item/2026-DON599}{\textit{Hantavirus cluster linked to cruise ship travel, Multi-country}}. Disease Outbreak News (4 May 2026).

\bibitem{supp-whonote0507}
World Health Organization. \href{https://www.who.int/news/item/07-05-2026-who-s-response-to-hantavirus-cases-linked-to-a-cruise-ship}{\textit{WHO's response to hantavirus cases linked to a cruise ship}}. Note for media (7 May 2026).

\bibitem{supp-whodon600}
World Health Organization. \href{https://www.who.int/emergencies/disease-outbreak-news/item/2026-DON600}{\textit{Hantavirus cluster linked to cruise ship travel, Multi-country}}. Disease Outbreak News (8 May 2026).

\bibitem{supp-whodg0512}
World Health Organization. \href{https://www.who.int/news-room/speeches/item/who-director-general-s-opening-remarks-at-the-media-briefing-on-hantavirus---12-may-2026}{\textit{WHO Director-General's opening remarks at the media briefing on hantavirus}}. Speech (12 May 2026).

\bibitem{supp-whodon601}
World Health Organization. \href{https://www.who.int/emergencies/disease-outbreak-news/item/2026-DON601}{\textit{Hantavirus cluster linked to cruise ship travel, Multi-country}}. Disease Outbreak News (13 May 2026).

\bibitem{supp-whodg0515}
World Health Organization. \href{https://www.who.int/news-room/speeches/item/who-director-general-s-opening-remarks-at-the-media-briefing---15-may-2026}{\textit{WHO Director-General's opening remarks at the media briefing}}. Speech (15 May 2026).

\bibitem{supp-si-whorra}
World Health Organization. \href{https://www.who.int/publications/m/item/who-rapid-risk-assessment---hantavirus-outbreak-caused-by-andes-virus--global-v.2}{\textit{Hantavirus outbreak caused by Andes virus, Global v.2}}. WHO Rapid Risk Assessment (17 May 2026).

\bibitem{supp-whodg0519}
World Health Organization. \href{https://www.who.int/news-room/speeches/item/who-director-general-s-address-to-member-states-at-the-79th-world-health-assembly---19-may-2026}{\textit{WHO Director-General's address to Member States at the 79th World Health Assembly}}. Speech (19 May 2026).

\bibitem{supp-whodg0522}
World Health Organization. \href{https://www.who.int/news-room/speeches/item/who-director-general-s-opening-remarks-at-the-member-state-information-session-on-outbreaks-of-ebola-and-hantavirus-22-may-2026}{\textit{WHO Director-General's opening remarks at the Member State information session on outbreaks of Ebola and hantavirus}}. Speech (22 May 2026).

\bibitem{supp-whohantavirus2026}
World Health Organization. \href{https://www.who.int/emergencies/disease-outbreak-news/item/2026-DON604}{\textit{Hantavirus outbreak linked to cruise ship travel, Multi-locations}}. Disease Outbreak News (28 May 2026).

\bibitem{supp-si-eurosurveillance}
van den Berg, O. E. et al. Andes virus outbreak linked to expedition cruise ship travel, multi-country investigation and response, April to June 2026. \href{https://doi.org/10.2807/1560-7917.ES.2026.31.24.2600477}{\textit{Euro Surveill.} \textbf{31}, 2600477 (2026)}.

\bibitem{supp-si-eurosurveillance-addendum}
Eurosurveillance editorial team. Addendum for Euro Surveill. 2026;31(24). \href{https://doi.org/10.2807/1560-7917.ES.2026.31.27.260709a}{\textit{Euro Surveill.} \textbf{31}, 260709a (2026)}.

\bibitem{supp-phac_dec2025}
Public Health Agency of Canada. \href{https://health-infobase.canada.ca/measles-rubella/archive/2025/}{\textit{Measles and Rubella Weekly Monitoring Report}}. Health Infobase (2025).

\bibitem{supp-nysdoh_ghu}
New York State Department of Health. \href{https://globalhealthreports.health.ny.gov/}{\textit{Global Health Update}}. Weekly reports (2025).

\bibitem{supp-paho_epiupdate}
Pan American Health Organization. \href{https://www.paho.org/en/documents/epidemiological-update-measles-americas-region-1-july-2025}{\textit{Epidemiological Update: Measles in the Americas Region}}. Epidemiological updates (2025).

\bibitem{supp-paho_bulletin}
Pan American Health Organization. \textit{Vigilancia del Sarampión, Rubéola y Síndrome de Rubéola Congénita en las Américas} (in Spanish). Biweekly bulletin (2025).

\bibitem{supp-whompox}
World Health Organization. \textit{Multi-country outbreak of mpox}. External situation reports nos 51--58 (2025).

\bibitem{supp-whoafro}
WHO Regional Office for Africa. \textit{Weekly Bulletin on Outbreaks and Other Emergencies}. Weekly bulletins (2025).

\end{thebibliography}
}
\end{document}